\documentclass[11pt]{article}
\usepackage{geometry}
\usepackage{graphicx}
\usepackage{subfigure}
\usepackage{amssymb}
\usepackage{amsmath}
\usepackage{epstopdf}
\usepackage{cite}
\usepackage{bbm}
\usepackage{multirow}
\usepackage{enumerate}
\usepackage{ulem}
\usepackage{xcolor}
\usepackage{appendix}
\usepackage[hyperindex=true,
          pdfstartview=FitH,
          bookmarksnumbered=true,
          bookmarksopen=true,
          citecolor=blue,
          linkcolor=blue,
          colorlinks=true,
          pdfborder=00]{hyperref}

\newcommand{\ba}{\begin{align}}
\newcommand{\ea}{\end{align}}
\newcommand{\be}{\begin{equation}}
\newcommand{\en}{\end{equation}}
\newcommand{\bea}{\begin{eqnarray}}
\newcommand{\ena}{\end{eqnarray}}

\usepackage{graphicx} % Required for inserting images

\title{Time-reversed Shannon entropy as a chaos indicator for non-integrable systems}

\author{Wenfu Cao$^1$\footnote{202411100001@stu.ujn.edu.cn},  Siyan Chen$^2$\footnote{chensiyan@zufe.edu.cn}, Hongsheng Zhang$^1$\footnote{sps\_zhanghs@ujn.edu.cn (corresponding author)}  \\
$^1$School of Physics and Technology, University of Jinan, \\ 336 West Road of Nan Xinzhuang, Jinan, Shandong 250022, China \\
$^2$School of Data Science, Zhejiang University of Finance and Economics, \\ No. 18 Xueyuan Street,  Xiasha Higher Education Zone, Hangzhou, Zhejiang 310018, China
}

\begin{document}

\maketitle

\begin{abstract}

 We propose a novel chaos indicator---time-reversed Shannon entropy (TRSE)---that leverages the interplay between time-reversal symmetry breaking and information entropy in curved spacetimes. By quantifying statistical discrepancies between forward and backward temporal evolution of particle orbits, TRSE robustly distinguishes chaotic from regular dynamics in non-integrable systems. In contrast, integrable systems exhibit stable, symmetric probability distributions preserved by conserved quantities such as the Carter constant. We validate the method through high-precision numerical simulations in both Kerr and Schwarzschild-Melvin black hole geometries, evolving trajectories forward and backward in time. Furthermore, we refine our previously introduced particle-pair mutual information (MIPP) and perform comprehensive parameter-space scans, revealing a strong quantitative agreement between MIPP and TRSE. The two indicators emerge as complementary probes of chaos: TRSE captures symmetry breaking in orbital evolution, while MIPP measures statistical correlations. Together, they establish a unified framework for diagnosing chaos in general relativistic systems, paving a new path to understand the fundamental nature of chaos in non-integrable systems.

 %We propose a novel physical concept of time-reversed Shannon entropy(TRSE) and establishes it as a chaos indicator for non-integrable systems in curved spacetimes.
%The key innovation lies in combining the fundamental principle of time-reversal symmetry breaking with information entropy, enabling the detection of chaos by quantifying statistical differences between the forward and backward temporal evolution of particle orbits. As a sharp contrast, integrable systems exhibit stable, symmetric probability distributions governed by conserved quantities, for example the Carter constant. The efficacy of TRSE is rigorously validated through numerical developments backwards and forwards in both Kerr spacetime and the
%Schwarzschild-Melvin black hole geometry. Furthermore, we study key optimizations to the particle pair mutual information(MIPP), which was proposed in our previous work.
%Parameter space scans confirm that the MIPP exhibits a high degree of consistency with the novel TRSE indicator. A high degree of consistency between the TRSE and MIPP approaches is demonstrated by a fine scan of the parameter space. Indeed, TRSE and MIPP constitute a pair of dual (complementary) schemes for detecting chaos, designed to probe orbital symmetry breaking and statistical correlations. We demonstrate that the TRSE approach paves a new path to understand the fundamental nature of chaos in non-integrable systems.

  \end{abstract}
%\keywords{Shannon entropy; Carter constant; Chaos}
 %\maketitle
\section{Introduction}
In the vast conceptual architecture of physics, entropy stands as a cornerstone---a profoundly universal and continuously evolving notion whose reach extends across centuries and disciplines. It offers a unifying lens through which the behavior of systems ranging from microscopic particles to the cosmos at large can be understood. The genesis of entropy lies in Carnot's analysis of ideal heat engines. Clausius later formalized the concept, explicitly introducing the term entropy  as a state function that quantifies the irreversibility of thermodynamic processes, whose articulation of the second law of thermodynamics---that the entropy of an isolated system never decreases---endows physical time with a preferred direction, thereby defining the so-called arrow of time \cite{Clausius:1865}. A profound microscopic interpretation was subsequently provided by Boltzmann, who established the statistical mechanical foundation of entropy through the celebrated formula, $S=KlnW$, which revealed a more fundamental statistical truth: entropy is proportional to the logarithm of the number of microstates corresponding to a given macrostate.
This revolutionary insight uncovered the statistical origin of the second law of thermodynamics and grounded the concept of disorder within a mathematical framework based on microscopic probabilities \cite{Boltzmann1897}.

In the 20th century, von Neumann extended the concept of entropy into quantum mechanics by defining the von Neumann entropy as a key measure of the purity of quantum states \cite{VonNeumann1932}. Inspired by this development, Shannon introduced the notion of information entropy in information theory, using it to quantify uncertainty or average information content \cite{Shannon1948}.
Shannon's entropy abstracts away from physical systems per se, enabling broad applications across diverse domains---particularly in information science and artificial intelligence---and laying the cornerstone for quantum information science.

This progressive generalization---from thermodynamic entropy to statistical, quantum, and informational forms---culminated in one of the most profound revolutions in theoretical physics during the 1970s. Bekenstein and Hawking demonstrated that black holes possess an intrinsic entropy proportional to the area of their event horizon, thereby establishing black holes as genuine thermodynamic entities \cite{Bekenstein:1973ur, Hawking:1974rv}.
Recent analyses of the GW250114 event have achieved decisive progress: the results not only confirm with high confidence the increase in the event horizon area after merger---directly verifying Hawking's area theorem and thus strongly supporting the black hole analogue of the second law of thermodynamics---but also identify the rotational properties of the final black hole through spectral analysis of the ringdown signal \cite{LIGOScientific:2025rid}. These findings advance  new frontiers in testing general relativity and probing spacetime structure in strong-field regimes.

Beyond its role in gravitational thermodynamics, the informational formulation of entropy---particularly Shannon entropy---has emerged as a powerful diagnostic tool for analyzing intrinsic randomness in dynamical systems.  In stochastic dynamical systems, trajectory-dependent entropy can be defined for individual stochastic trajectories, leading to the proof of a universal fluctuation theorem: $\langle e^{-\Delta s_{tot}} \rangle=1$ \cite{Seifert:2005rlb, Leggio:2013wrr}.
Recently, Shannon entropy has been creatively applied to probe chaos in dynamical systems within curved spacetime \cite{Cao:2024bjk, Cao:2024rvo}.
In curved spacetime, particle orbits around black holes are prone to chaotic behavior due to external perturbations \cite{Wald:1974np, Karas:1992, Li:2018wtz, Cao:2024ihv,
Xu:2024ble, Li:2025jfq}.
By computing the Shannon entropy of orbital data, one can quantify the degree of uncertainty in motion: periodic orbits exhibit stable entropy values with small fluctuations, whereas chaotic orbits display significant entropy fluctuations. This dynamical entropy fluctuation resonates conceptually with thermodynamic fluctuation theorems, suggesting a deep structural link between deterministic chaos and statistical physics.

To improve both the robustness and efficiency in chaos detection, the notion of pairwise mutual information has been introduced. Based on the information-theoretic notion of mutual information, this measure quantifies the statistical dependence between two nearby orbital trajectories: regular orbits, characterized by strong trajectory correlations, yield pairwise mutual information values close to unity, while chaotic orbits, exhibiting exponential divergence, result in values approaching zeros. Compared to the Fast Lyapunov Indicator (FLI), pairwise mutual information demonstrates superior consistency in parameter scans and higher computational efficiency, without requiring renormalization procedures.

Building on several successes achieved in our previous works, challenges persist in applying Shannon entropy to identify chaos in complex dynamical systems.  First, it struggles to distinguish true chaos from complex quasiperiodic motion, as both may produce similarly high entropy values over the finite observation times. Second, from a methodological standpoint, the computation of entropy fluctuations relies on partitioning orbital data into discrete bins, where the horizontal axis corresponds to bin indices instead of physical time---an approach that introduces a certain degree of arbitrariness. While this method captures entropy fluctuations, it obscures the temporal continuity of orbital evolution, thereby limiting our ability to extract fine-grained physical information from the dynamical process. To address these limitations, we propose and explore a global entropy calculation method---one that computes a single entropy value for an entire trajectory, aiming to transcend purely statistical interpretations and more directly reveal the entropic evolution law of individual orbits in a more physically clear and  coherent manner.

This article is organized as follows. In section \ref{sec:two} we introduce the concepts of the Killing tensor and Carter constant, analyze the probability distribution of particle orbits based on these symmetries, and review the characteristics of Shannon entropy and pairwise mutual information methods. In section \ref{sec:three} we develop and apply the time-reversed Shannon entropy method to probe chaotic motion: first in the Kerr spacetime, then in the more complex Schwarzschild-Melvin geometry, focusing on the transition of non-chaotic/chaotic behaviors of photon orbits. In section \ref{sec:four} we summarize our results, discuss their implications, and outline potential directions for future research.
\section{Background: from orbits to chaos indicators}\label{sec:two}
In this section, we first introduce the concept of hidden symmetry in the spacetime of general relativity, then explain its control over the orbital distribution of particles in the polar
angle ($\theta$) direction, and provide a review of existing chaos indicators based on Shannon entropy.
\subsection{ Separability of  Hamilton-Jacobi system}
The Hamilton-Jacobi(HJ) equation for a test particle in curved spacetime is given by
\begin{eqnarray}
\frac{\partial S}{\partial \tau}+\frac{1}{2}g^{ab}\frac{\partial S}{\partial x^{a}}\frac{\partial S}{\partial x^{b}}&=&0,
\end{eqnarray}
based on the symmetry of spacetime, the solution to the HJ equation is assumed to have a separable form
\begin{eqnarray}
S&=&\frac{1}{2}\mu^{2}\tau-Et+L_{z}\phi+S_{r}(r)+S_{\theta}(\theta),
\end{eqnarray}
where $E$ and $L_{z}$ are conserved quantities arising from the Killing vector fields, representing the energy and angular momentum of the test particle, respectively.
$\mu$ is the rest mass of the test particle.
The functions $S_{r}(r)$ and $S_{\theta}(\theta)$ are yet to be determined, which are governed by the hidden symmetries induced by the Killing tensor fields.
According to the Benenti-Francaviglia theorem \cite{Benenti:1979erw, Papadopoulos:2018nvd}, in Boyer-Lindquist type coordinates$(t,r,x,\phi)$, the HJ equation of spacetime locally has a separable
structure if and only if the spacetime possesses two Killing vector fields and two Killing tensor fields. The two Killing vector fields are well established.
Among the two Killing tensor fields, one is the trivial metric tensor $g^{ab}$, which corresponds to the conservation of the particle's rest mass, while the other is non-trivial.
It is the presence of this non-trivial Killing tensor that directly determines the separability of the HJ equation and gives rise to a fourth constant of motion.
Therefore, spacetimes that satisfy the above separability conditions must take a specific form for their contravariant metric tensor in Boyer-Lindquist type
coordinates$(x=\cos\theta)$\cite{Papadopoulos:2018nvd}
\begin{eqnarray}
g^{\mu\nu} \partial_{\mu} \partial_{\nu}&=&\frac{1}{A_{1}(r)+B_{1}(x)}[(A_{5}(r)+B_{5}(x))(\partial_{t})^{2}+A_{2}(r)(\partial_{r})^{2}+B_{2}(x)(\partial_{x})^{2} \nonumber
\\&&  +(A_{3}(r)+B_{3}(x))(\partial_{\phi})^{2}+2(A_{4}(r)+B_{4}(x))\partial_{t}\partial_{\phi}],
\end{eqnarray}
while the corresponding non-trivial contravariant Killing tensor is
\begin{eqnarray}
K^{\mu\nu} \partial_{\mu} \partial_{\nu}&=&\frac{1}{A_{1}(r)+B_{1}(x)}[(B_{1}(x)A_{5}(r)-A_{1}(r)B_{5}(x))(\partial_{t})^{2}+A_{2}(r)B_{1}(x)(\partial_{r})^{2}-B_{2}(x)A_{1}(r)(\partial_{x})^{2}
\nonumber
\\&&  +(B_{1}(x)A_{3}(r)-A_{1}(r)B_{3}(x))(\partial_{\phi})^{2}+2(B_{1}(x)A_{4}(r)-A_{1}(r)B_{4}(x))\partial_{t}\partial_{\phi}],
\end{eqnarray}
and satisfies $\nabla_{(c}K_{ab)}=0$.
The Hamiltonian of the test particle($H=\frac{1}{2}g^{ab}p_{a}p_{b}$), together with Eq. (3), provides the on-shell condition
\begin{eqnarray}
0 & = & \frac{A_5 E^2}{A_1 + B_1} + \frac{B_5 E^2}{A_1 + B_1} - \frac{2 A_4 E L_z}{A_1 + B_1} - \frac{2 B_4 E L_z}{A_1 + B_1} + \frac{A_3 L_z^2}{A_1 + B_1} \nonumber \\
& & + \, \frac{B_3 L_z^2}{A_1 + B_1} + \frac{A_2 p_r^2}{A_1 + B_1} + \frac{B_2 p_x^2}{A_1 + B_1} +\mu^2.
\end{eqnarray}

The Carter-like constant $K$ is given by
\begin{eqnarray}
&&K=K_{ab}\dot{x}^{a}\dot{x}^{b}=K^{ab}p_{a}p_{b}= \frac{A_5 B_1 E^2}{A_1 + B_1} - \frac{A_1 B_5 E^2}{A_1 + B_1} - \frac{2 A_4 B_1 E L_z}{A_1 + B_1} + \frac{2 A_1 B_4 E L_z}{A_1 + B_1} \nonumber
\\
&& \quad + \frac{A_3 B_1 L_z^2}{A_1 + B_1} - \frac{A_1 B_3 L_z^2}{A_1 + B_1} + \frac{A_2 B_1 p_r^2}{A_1 + B_1} - \frac{A_1 B_2 p_x^2}{A_1 + B_1}.
\end{eqnarray}

Using Eq. (5), one can simplify Eq. (6) as
\begin{eqnarray}
K&=& B_5 E^2 - 2 B_4 E L_z + B_3 L_z^2 + B_2 p_x^2 + B_1\mu^{2},
\end{eqnarray}
or equivalently as
\begin{eqnarray}
K&=& -A_5 E^2 + 2 A_4 E L_z - A_3 L_z^2 - A_1 \mu^2 - A_2 p_r^2.
\end{eqnarray}
These four constants of motion $(E, L_{z}, K, \mu)$ commute with each other. According to the Liouville's theorem, this means that the motion of the particle is confined to a 4-dimensional submanifold
within an 8-dimensional phase space, thereby making the geodesic motion integrable \cite{Masoliver:2011}.

By combining  $\dot{t}=g^{tt}p_{t}+g^{t\phi}p_{\phi}$, $\dot{\phi}=g^{t\phi}p_{t}+g^{\phi\phi}p_{\phi}$, as well as Eqs. (3), (7), and(8), one can derive the first-order equations of motion for
the
test particle:
\begin{eqnarray}
(A_1 + B_1)\frac{dt}{d\tau}&=& -(A_5 + B_5) E + (A_4 + B_4) L_z,
\nonumber\\ \nonumber
(A_1 + B_1)\frac{dr}{d\tau}&=& \pm\sqrt{-A_2A_5 E^2 +2A_2A_4 E L_z-A_2A_3 L_z^2 -A_2(A_1\mu^2+K)},
\\
(A_1 + B_1)\frac{d\theta}{d\tau}&=& \pm\sqrt{K-B_5 E^2 + 2 B_4 E L_z - B_3 L_z^2- B_1\mu^{2}} ,
\\ \nonumber
(A_1 + B_1)\frac{d\phi}{d\tau}&=& -(A_4 + B_4) E+ (A_3 + B_3) L_z .
\end{eqnarray}

We use the 8th-order Runge-Kutta method to numerically integrate the system of Eq. (9).
The orbits discussed in this paper all satisfy the condition for bound orbits($E^{2}\leq1$) \cite{1972PhRvD...5..814W, 1972ApJ...178..347B, Teo:2020sey, 2024ApJ...966..226K}.
The accuracy and stability of the numerical integration are confirmed by monitoring energy conservation, with the relative energy error maintained at the order of $10^{-10}$ over the entire evolution.
\subsection{Probability distribution of order and chaotic orbits}
We begin our analysis with the Schwarzschild black hole.
Its spherical symmetry and the consequent integrability of particle orbits serve as an ideal starting point, which provides the clearest possible framework for analyzing orbital probability
distributions.
The needed allocations are
\begin{eqnarray}
&& A_{1}(r)=r^{2}, A_{2}(r)=r^{2}, A_{3}(r)=0, A_{4}(r)=0, A_{5}(r)=-r^{2},\nonumber
\\
&& B_{1}(x)=0, B_{2}(x)=\sin^{2}\theta, B_{3}(x)=\frac{1}{\sin^{2}\theta}, B_{4}(x)=0, B_{5}(x)=0.
\end{eqnarray}

Inserting Eq. (10) into Eq. (7) yields
\begin{eqnarray}
\Theta(\theta)=Q-\cos^{2}\theta\frac{L_{z}^{2}}{\sin^{2}\theta},S_{\theta}(\theta)=\pm\int d\theta\sqrt{\Theta(\theta)}~,
\end{eqnarray}
where  Carter constant Q relates with the Carter-like constant K through $ Q=K-(L_{z}-aE)^{2}$ \cite{Carter:1968rr}.
It is worth noting that the function $\Theta(\theta)$ possesses an intrinsic symmetry within the interval $(0,\pi)$, meaning it remains invariant under the coordinate transformation
$\theta\rightarrow\pi-\theta$.
From a physical perspective, this transformation is equivalent to a reflection symmetry with respect to the equatorial plane.
This property leads to the probability density function (PDF) $f_{\theta}(\theta)$ also being symmetric about the equatorial plane,
implying that the orbital behavior in the northern and southern hemispheres is statistically indistinguishable.
In numerical integration, due to the cumulative changes in the coordinate $\theta$, its values may exceed the interval $(0,\pi)$.
Therefore, using a bounded variable $x\equiv \cos\theta(-1<x<1)$ to replace the coordinate $\theta$ is useful.

Assume that the PDF $f_{\theta}(\theta)$ is proportional to $\frac{1}{p_{\theta}}$, then the PDF $f_{x}(x)$ is given by
\begin{eqnarray}
&& f_{x}(x)=f_{\theta}(\theta) |\frac{d\theta}{dx}|\propto \frac{1}{p_{\theta}} |\frac{d\theta}{dx}|\propto \frac{1}{\sqrt{Q(1-x^{2})-L_{z}^{2}x^{2}}}.
\end{eqnarray}
Although Eq. (12) becomes singular at the boundaries, it retains qualitative value since it effectively captures the essential feature of the probability
distribution for orbits along the polar angle $\theta$. To validate the theoretical prediction, we perform a numerical analysis of orbital motions.

The effective potential is critical for analyzing circular orbits in Schwarzschild spacetime.
By examining its extremum points, one can efficiently determine the orbital radii and assess their stability conditions, such as identifying the  stable circular orbit (SCO).
The effective potential for particles moving in the plane defined by the angle  $\theta$ is \cite{Cao:2024ihv}
\begin{eqnarray}
&& V_{eff}=E^{2}=(1+\frac{K}{r^{2}})(1-\frac{2}{r}), \nonumber\\
&& K=L^{2}_{z}\csc^{2}\theta.
\end{eqnarray}

 \begin{figure*}[h]
\begin{center}
\includegraphics[width=0.43 \textwidth]{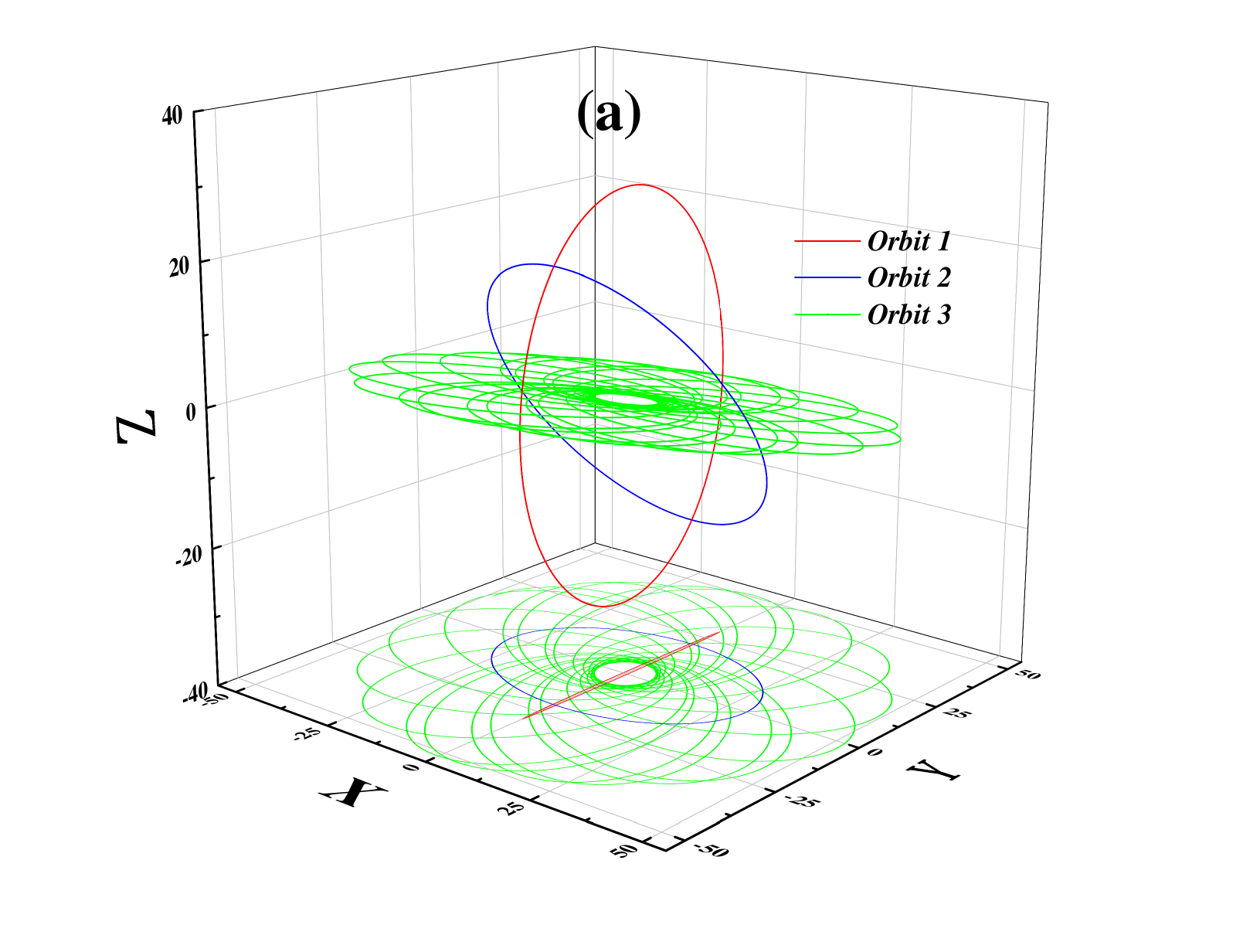}
\includegraphics[width=0.43 \textwidth]{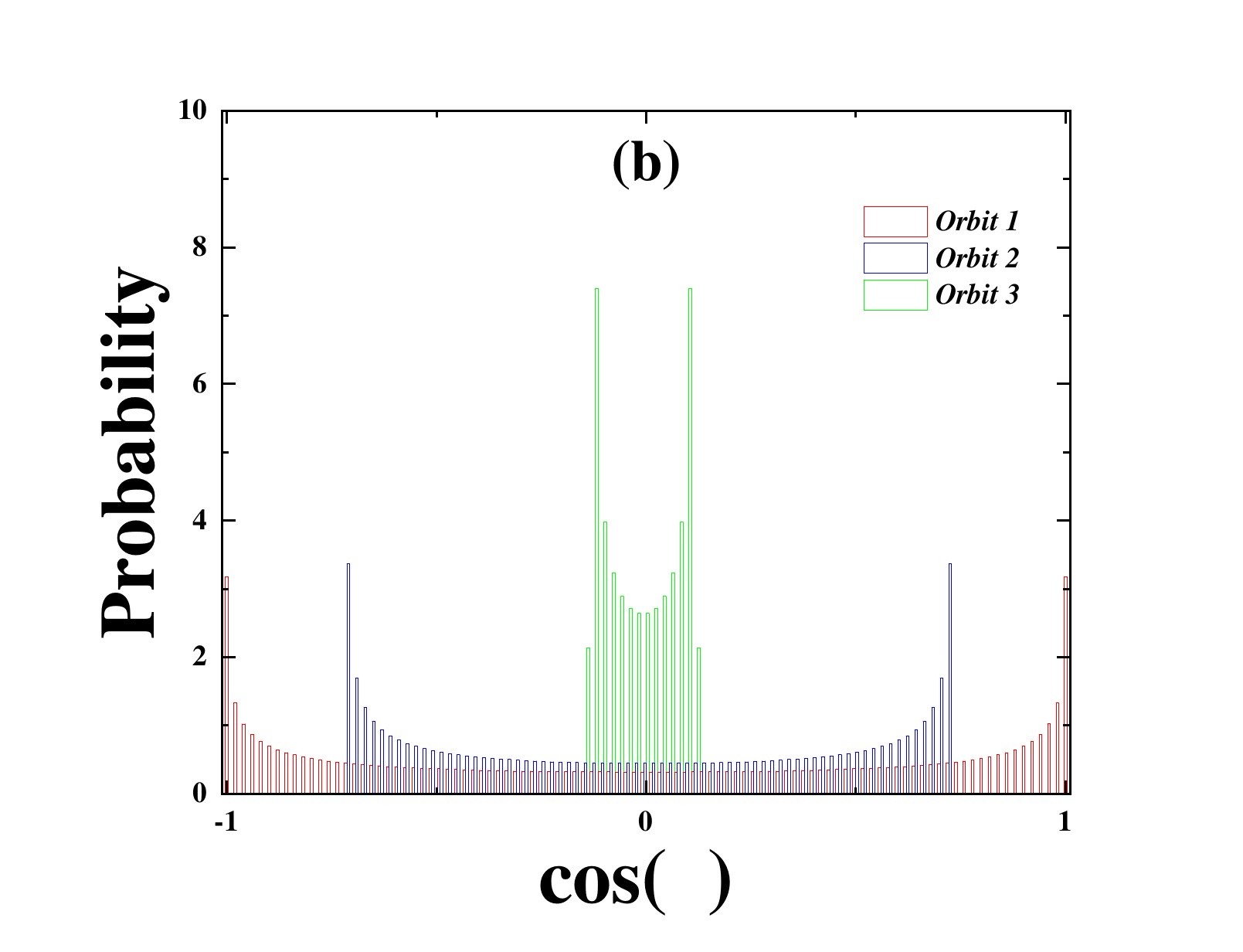}
\includegraphics[width=0.43 \textwidth]{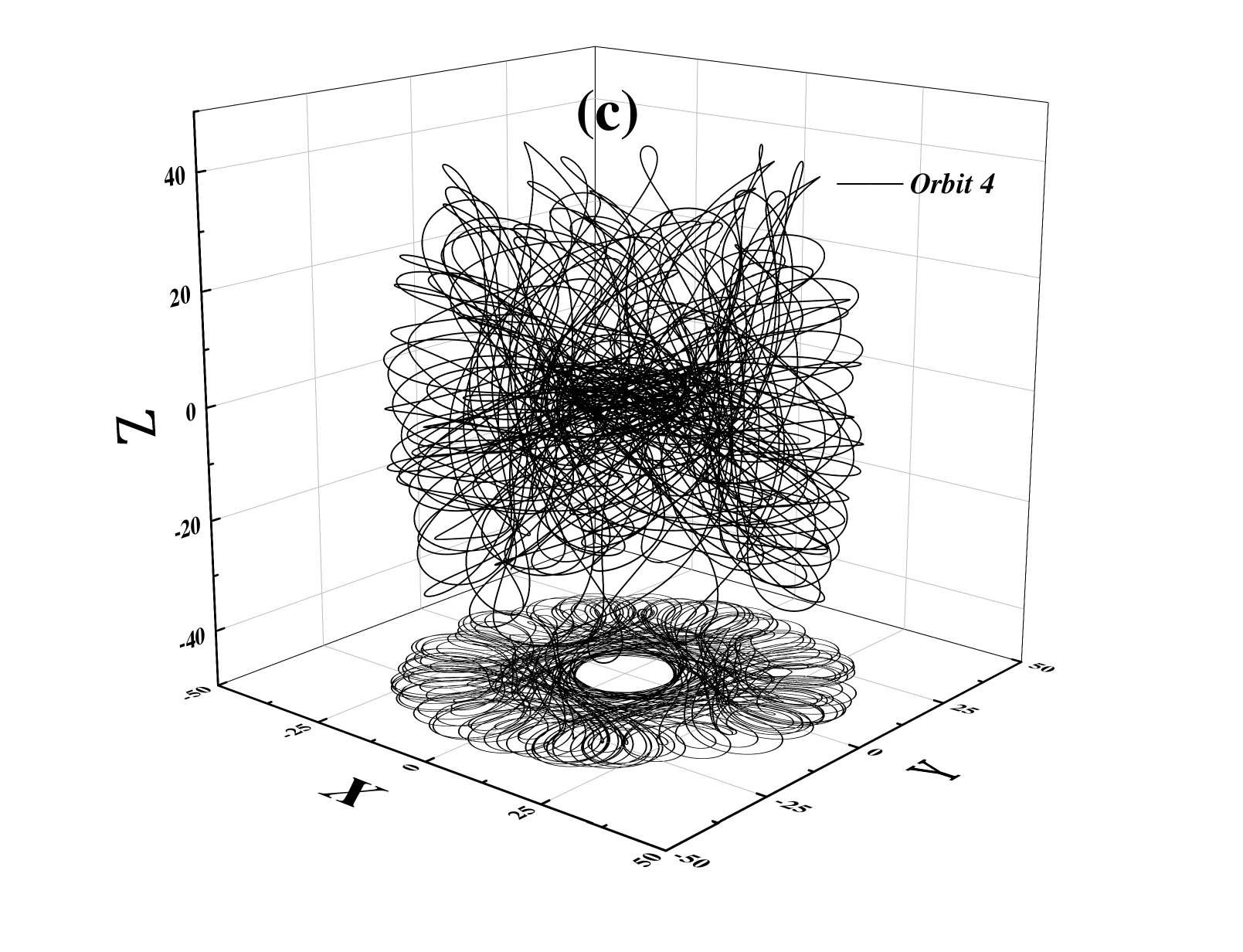}
\includegraphics[width=0.43 \textwidth]{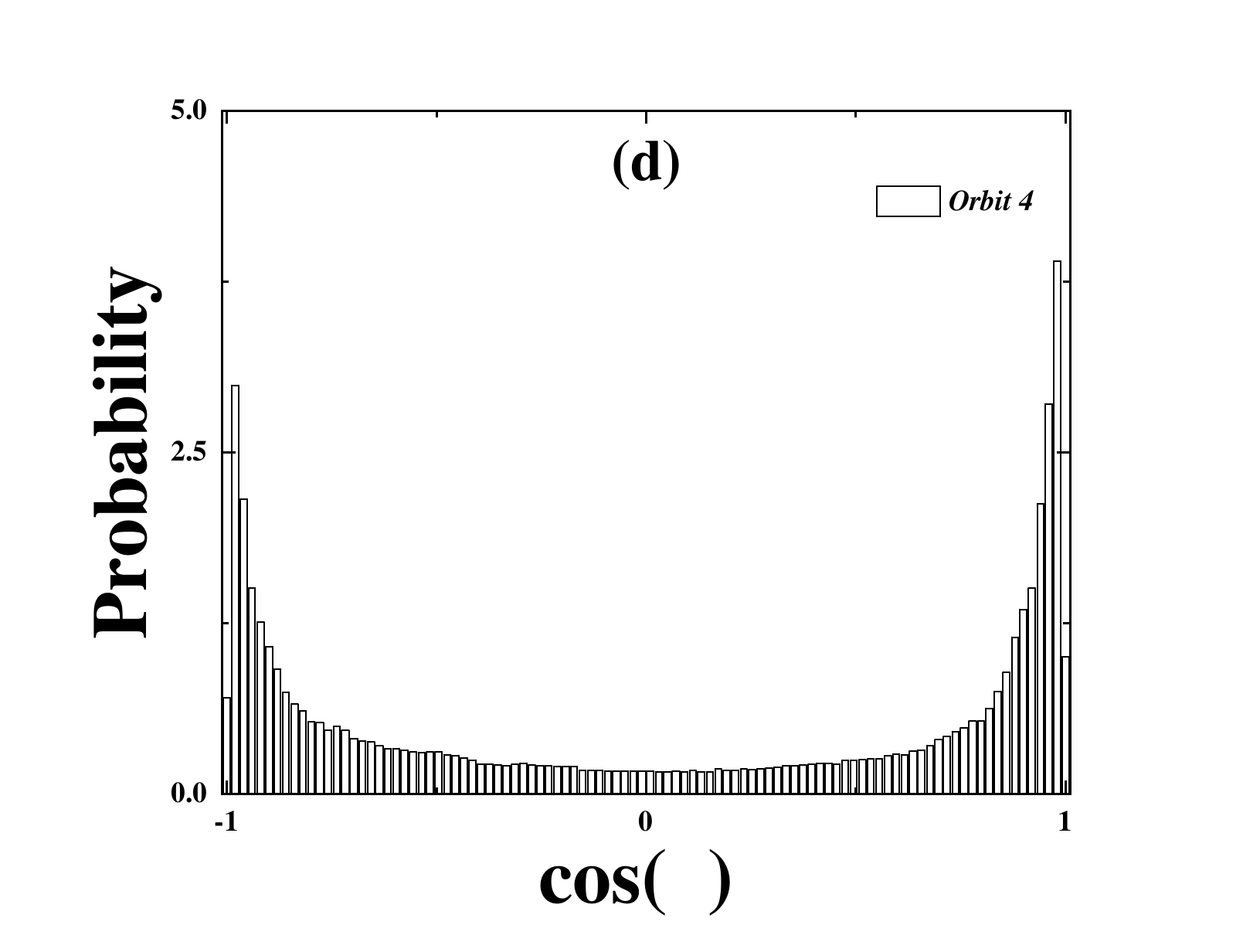}
\end{center}
\caption{Orbital trajectories and probability distributions in Schwarzschild spacetime.
(a) and (b) The system parameters for Orbit 1 are $E=0.983$, $L_{z}=0.1$, $r=29.49$, and $\theta=1^{\circ}$.
The system parameters for Orbit 2 are $E=0.983$, $L_{z}=4$, $r=29.49$, and $\theta=90^{\circ}$.
The system parameters for Orbit 3 are $E=0.983$, $L_{z}=4$, $r=6$, and $\theta=90^{\circ}$.
(c) and (d) The system parameters for Orbit 4 are $E=0.98$, $L_{z}=4.6$, B=0.01, $r=16$, and $\theta=90^{\circ}$.}
 \label{Fig1}
\end{figure*}

As shown in Fig. 1(a), three order orbits with different characteristics are obtained through the effective potential method. Orbit 1 (red) possesses minimal angular momentum, and its orbital plane closely
aligned to the polar axis.
The particle covers nearly all latitudes from the North Pole to the South Pole during its orbit, but its velocity is slower in the high-latitude regions, and the residence time
is significantly longer.
Orbit 2 (blue), due to an increase in angular momentum and initial release at the equatorial plane, forms an elliptical orbit oscillating near the equator.
Orbit 3 (green) exhibits a regular precessing orbit.
As shown in Fig 1(b), the probability density functions (PDFs) of these three types of orbits all exhibit a strict symmetric distribution about the equatorial plane. This symmetry is highly consistent with the theoretical
formula,
and its physical origin lies in the dynamical symmetry guaranteed by the Carter constant.

When the system experiences an external magnetic field perturbation, it undergoes a fundamental change. When the magnetic field strength exceeds a certain threshold, the particle
motion exhibits extreme sensitivity to initial conditions, leading to the onset of chaotic behavior, which then dominates. At this point, the method of separation of variables in the HJ equation
is
no longer applicable, and it becomes impossible to obtain an analytical solution for the orbital motion. To study this complex dynamics, we must instead  numerically solve the Hamiltonian
canonical
equations \cite{Sun:2021oxg}
\begin{eqnarray}
&& H=\frac{1}{2}g^{\mu\nu}(p_{u}-qA_{\mu})(p_{\nu}-qA_{\nu}), \nonumber\\
&& \dot{x}^{\mu}=\frac{\partial H}{\partial p_{\mu}}, \dot{p}_{u}=-\frac{\partial H}{\partial\dot{x}^{\mu}},
\end{eqnarray}
where $A_{\mu}$ is the electromagnetic potential.
Fig. 1(c) shows a chaotic orbit (Orbit 4) generated under the influence of a magnetic field perturbation.
Its three-dimensional trajectory exhibits a highly complex, entangled structure, which folds and diffuses randomly in space, filling the entire plotting area.
Fig. 1(d) further reveals the statistical properties of the PDF of this orbit.
While the overall distribution still roughly exhibits a "U"-shaped symmetry about the equatorial plane, reflecting the system's residual macroscopic symmetry, the PDF shows sharp and irregular
fluctuations at a local scale, especially near the equatorial plane.
These local fluctuations are a direct manifestation of the intrinsic randomness of the chaotic system.
This phenomenon indicates that the loss of the Carter constant breaks the dynamical symmetry, leading to the motion of the orbit in the polar direction losing its strict constraints.
The statistical result no longer reflects the "perfect symmetry" guaranteed by conserved quantities in regular orbits, but instead degenerates into a mixed state where macroscopic approximate
symmetry coexists with microscopic random fluctuations.
\subsection{Comments on Shannon entropy and MIPP}
In the dynamics of curved spacetime, the reliability and efficiency of chaos detection methods are core issues.
In the scan plot of the magnetic field parameter $B$, as shown in Fig. 2(a), the Shannon entropy value of the blue orbit (regular) is higher than that of the red orbit (chaotic).
This indicates that the entropy value alone is not sufficient to reliably distinguish between the regular and chaotic states of the orbits.
However, recent studies have shown that the fluctuation characteristics of the entropy value time series can serve as a sensitive indicator for identifying chaos.
As shown in Fig. 2(c), the Shannon entropy of the chaotic orbit (red) exhibits significantly greater fluctuations compared to the regular orbit (blue).
The advantage of this method lies in its visual assessment, but its effectiveness is highly dependent on the correct construction of the probability distribution.
In view of such situations, we propose a significantly refined approach, in which we utilize the periodicity and boundaries of the polar angle $(\theta)$ to
construct the probability distribution, to replace the conventional approach of arbitrarily setting the radial coordinate range.

This approach inherently captures the statistical nature of orbital motion, thereby improving the reliability of the results.
The cost, however, in order to accurately analyze the fluctuation characteristics, the orbit must be divided into multiple intervals for statistical analysis, which may incur a higher computational power \cite{Cao:2024bjk}.

%\begin{figure*}[htbp]
%\center{
%\includegraphics[scale=0.2]{fig2a.pdf}
%\includegraphics[scale=0.2]{fig2b.pdf}
%\includegraphics[scale=0.2]{fig2c.pdf}
%\includegraphics[scale=0.2]{fig2d.pdf}
%\caption{ MIPP and Shannon entropy scanplots for the magnetic field parameter B in Schwarzschild spacetime. The range of the magnetic field parameter B is from 0.0002 to 0.01,
%and a total of 50 orbits were calculated. The system parameters are $E=0.99$, $L_{z}=4$, $r=16$, and $\theta=90^{\circ}$.}
 %\label{Fig1}}
%\end{figure*}

 \begin{figure*}[h]
\begin{center}
\includegraphics[width=0.43  \textwidth]{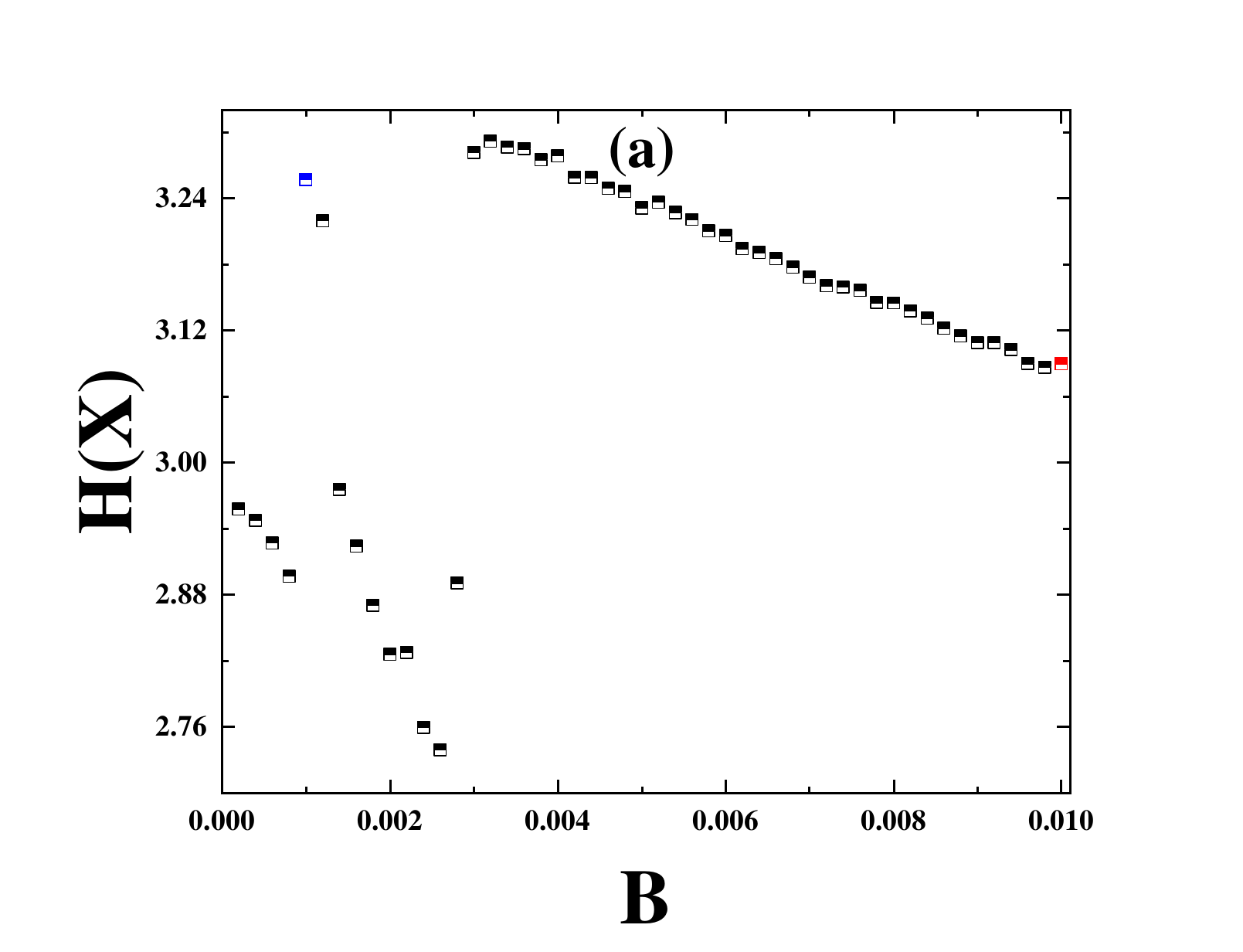}
\includegraphics[width=0.43  \textwidth]{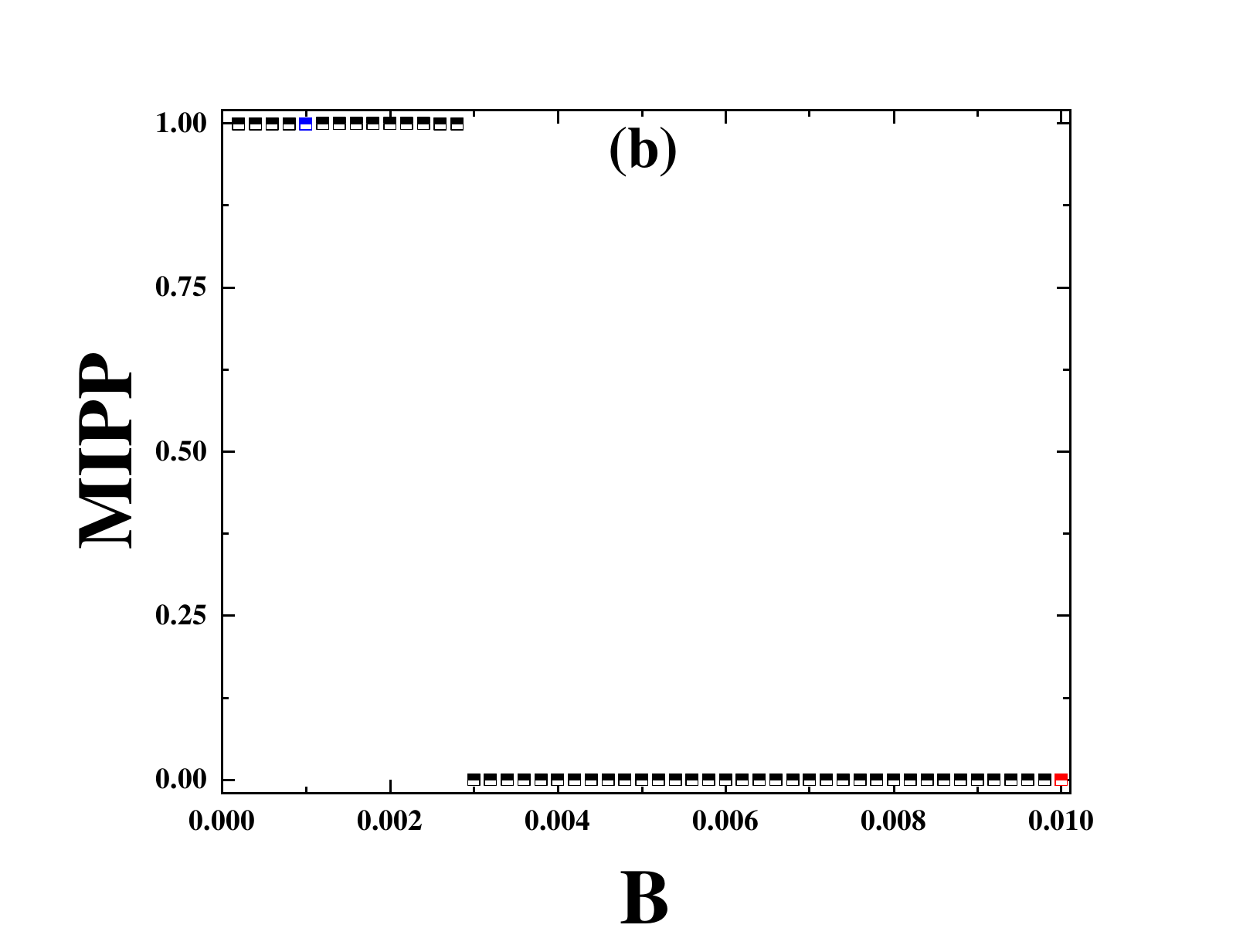}
\includegraphics[width=0.43  \textwidth]{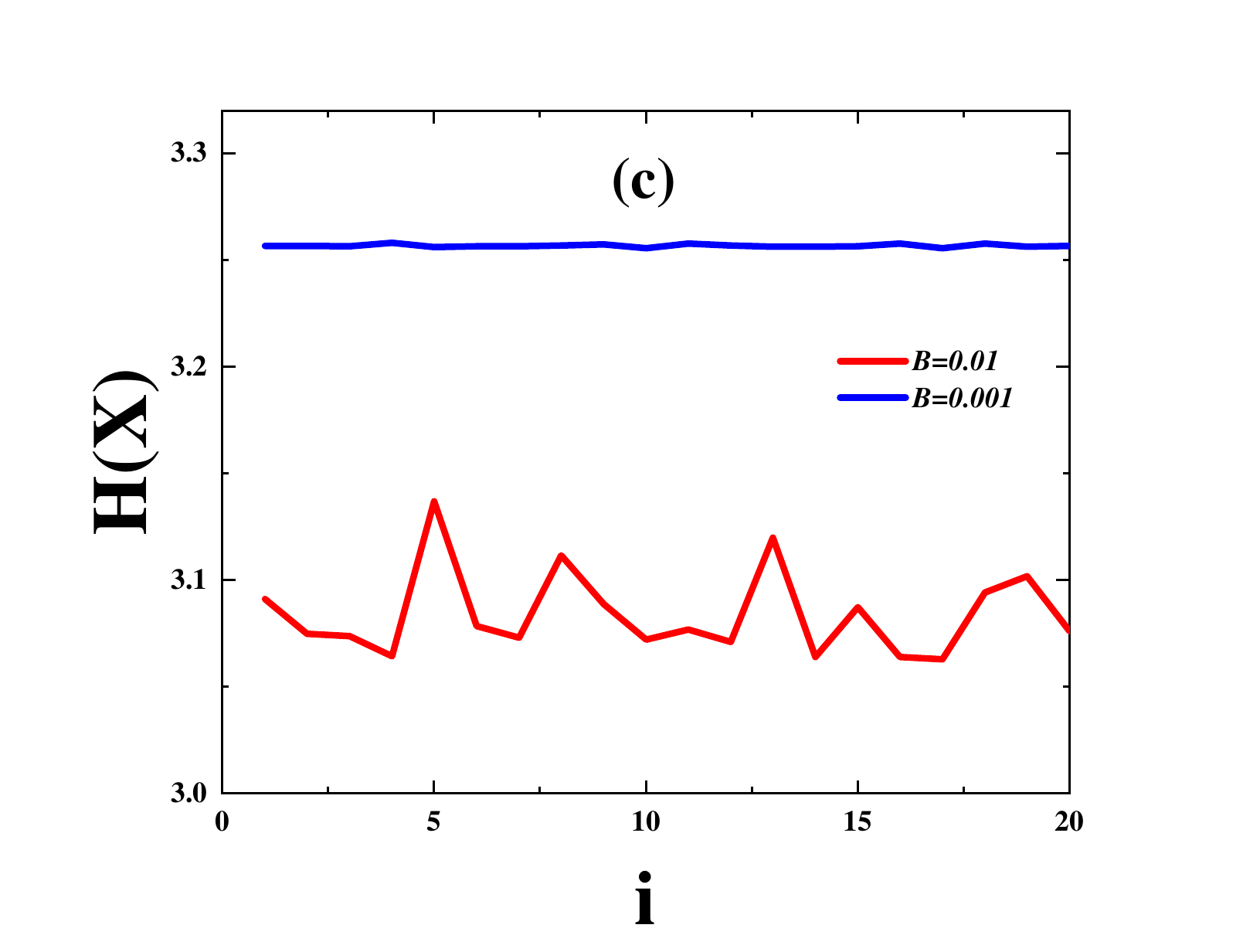}
\includegraphics[width=0.43  \textwidth]{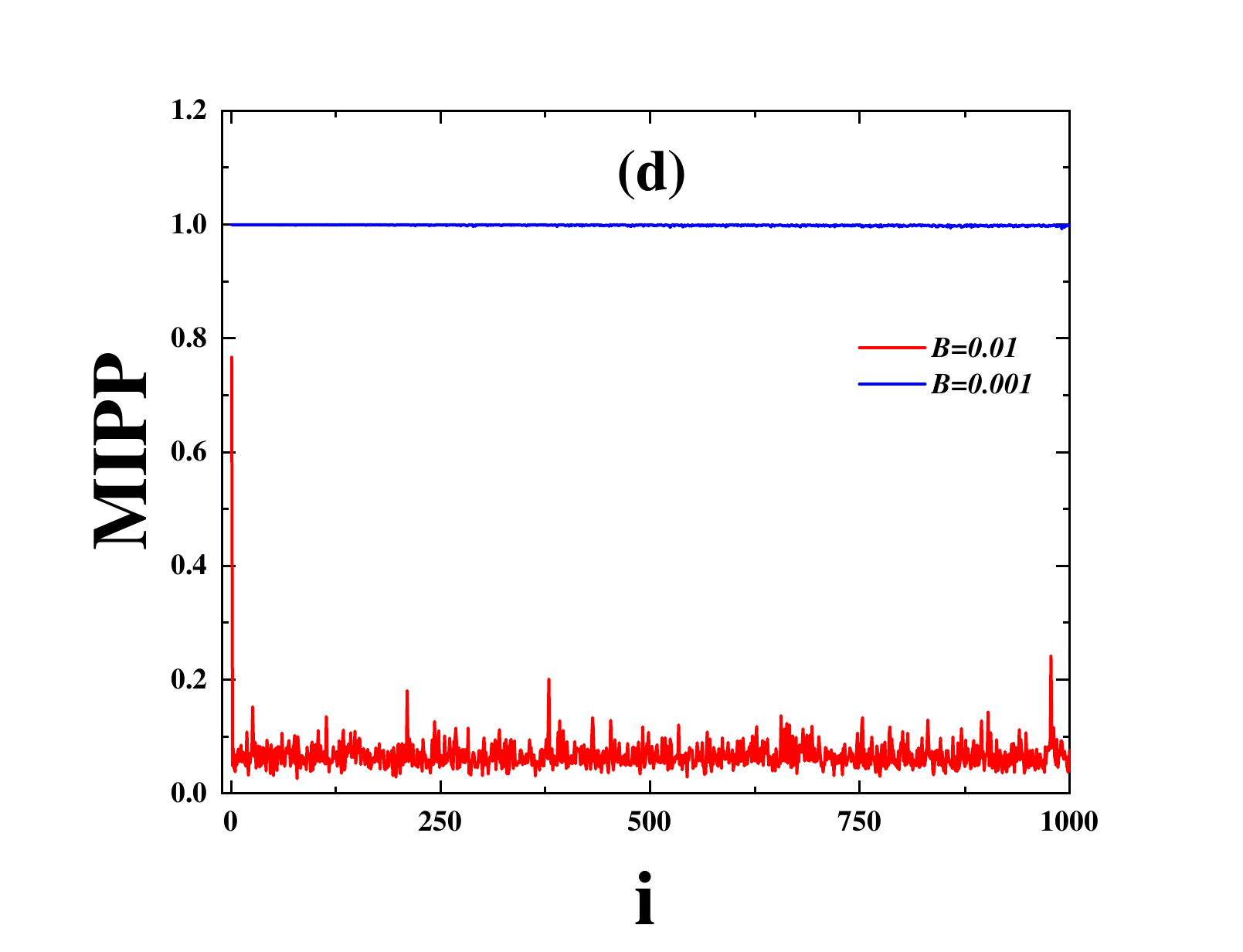}
\end{center}
\caption{ MIPP and Shannon entropy scanplots for the magnetic field parameter B in Schwarzschild spacetime. The range of the magnetic field parameter B is from 0.0002 to 0.01,
and a total of 50 orbits were calculated. The system parameters are $E=0.99$, $L_{z}=4$, $r=16$, and $\theta=90^{\circ}$.}
 \label{Fig2}
\end{figure*}

In contrast, the MIPP method provides more efficient diagnostics by quantifying the statistical dependence between the trajectories of neighboring orbits.
Its computational process is flexible: as shown in Fig. 2(b), it can perform a global calculation to quickly obtain an overall criterion, or, as shown in Fig. 2(d), it can perform interval-based
calculations to reveal the dynamic evolution of the MIPP values.
The latter clearly indicates that the MIPP value of the chaotic orbit rapidly decays from 1 and fluctuates around 0, rather than remaining strictly zero.
This reveals that there exists a weak but dynamic statistical correlation between the chaotic orbits, indicating that chaos is not merely a random disorder but a complex dynamic process with an inherent structure\cite{Cao:2024rvo}.
\section{The time-reversal Shannon entropy}\label{sec:three}
As established in Section II, the traditional Shannon entropy is fundamentally limited in distinguishing  between regular and chaotic motion.
To address this limitation, we introduce a new chaos indicator: TRSE.
This section specifically presents the theoretical formulation of this indicator.
We then validate its efficacy in the Kerr spacetime. Finally, we apply TRSE to the  complex Schwarzschild-Melvin black hole model to detect chaotic photon orbits.
\subsection{Formulation of the Time-reversal Shannon entropy indicator}
The theoretical derivation of the TRSE method is based on the assumption of small perturbations in the dynamical evolution.
We consider an idealized undisturbed dynamical system, where the probability density function of the system's orbit is denotes as $f(x)$.
Although the system's dynamical equations have time-reversal symmetry mathematically, rounding errors in numerical calculations cause the time-reversed evolution of chaotic orbits to deviate
exponentially. As a result, the entropy difference for chaotic orbits becomes more significant than for order orbits.
Based on this, we assume that the PDFs of the particle under forward and time-reversed evolution are respectively \cite{Seifert:2005rlb}
\begin{eqnarray}
&& p_{1}(x)=f(x)+g(x),\\
&& p_{2}(x)=f(x)+h(x),
\end{eqnarray}
in this framework, through mathematical derivation, we can obtain the approximate expression for the entropy difference
\begin{eqnarray}
\Delta H=-\int(\ln f(x)+1)(g(x)-h(x))dx-\frac{1}{2}\int\frac{1}{f(x)}(g(x)^{2}-h(x)^{2})dx+O(\|g(x)\|^{3}).
\end{eqnarray}
In the discrete form, suppose we divide the sampling space into $N$ discrete grids.
Since the sum of probabilities must equal 1, this means that the total sum of perturbation terms must be zero
\begin{eqnarray}
\sum_{i=1}^{N} g_i = 0, \quad \sum_{i=1}^{N} h_i = 0,
\end{eqnarray}
therefore, the entropy difference in discrete form can be further expressed as
\begin{eqnarray}
\Delta H\approx-\sum_{i=1}^{N}(g_i-h_i)\ln f_{i}-\frac{1}{2}\sum_{i=1}^{N}\frac{g^{2}_{i}-h^{2}_{i}}{f_{i}},
\end{eqnarray}
the first-order term in Eq. (19) is linearly dependent on the perturbation difference between forward and reverse evolutions, directly reflecting the degree of symmetry breaking in the system under numerical evolution.
The second-order term captures the nonlinear square difference of the perturbation amplitudes.
The exponential sensitivity of chaotic orbits to initial conditions causes the perturbation functions $g(x)$ and $h(x)$ to exhibit significant differences, whereas the stability of regular orbits
results in $g\approx h$, leading to $\Delta H\approx0$.
In the subsequent calculations, to calculate the TRSE, we numerically integrate each orbit both forward (from $t\rightarrow T$) and in reverse (from $T\rightarrow t$).
The integration time for both cases is $T=10^{7}$ to ensure statistical sufficiency.
Next, we extract the orbital polar angle $\theta$ and convert it to the variable $x=\cos\theta$.
We then construct the probability distribution for the forward and backward evolutions, respectively.
Ultimately, to enhance the comparability of entropy differences between different orbits, we  adopt the signed logarithmic scale $\pm log_{10}|\Delta H|$ as a chaos indicator,
where a positive sign indicates that the complexity of the time-reversed orbit is lower than that of the forward orbit, and a negative sign indicates that the complexity of the forward orbit is
lower than that of the time-reversed orbit.
\subsection{Validation in Kerr Spacetime}
The needed allocations for Kerr spacetime are:
\begin{eqnarray}
&& A_{1}(r)=r^{2}, A_{2}(r)=\Delta, A_{3}(r)=-\frac{a^{2}}{\Delta}, A_{4}(r)=-\frac{r^{2}+a^{2}}{\Delta}a, A_{5}(r)=-\frac{(r^{2}+a^{2})^{2}}{\Delta},\nonumber
\\
&& B_{1}(x)=a^{2}\cos^{2}\theta, B_{2}(x)=\sin^{2}\theta, B_{3}(x)=\frac{1}{\sin^{2}\theta}, B_{4}(x)=a, B_{5}(x)=a^{2}\sin^{2}\theta,
\end{eqnarray}
where $\Delta=r^{2}+a^{2}$. Inserting Eq. (18) into Eq. (7) yields
\begin{eqnarray}
\Theta(\theta)=Q-\cos^{2}\theta[a^{2}(\mu^{2}-E^{2})+\frac{L_{z}^{2}}{\sin^{2}\theta}],S_{\theta}(\theta)=\pm\int d\theta\sqrt{\Theta(\theta)}.
\end{eqnarray}

\begin{figure*}[htbp]
\center{
\includegraphics[width=0.43  \textwidth]{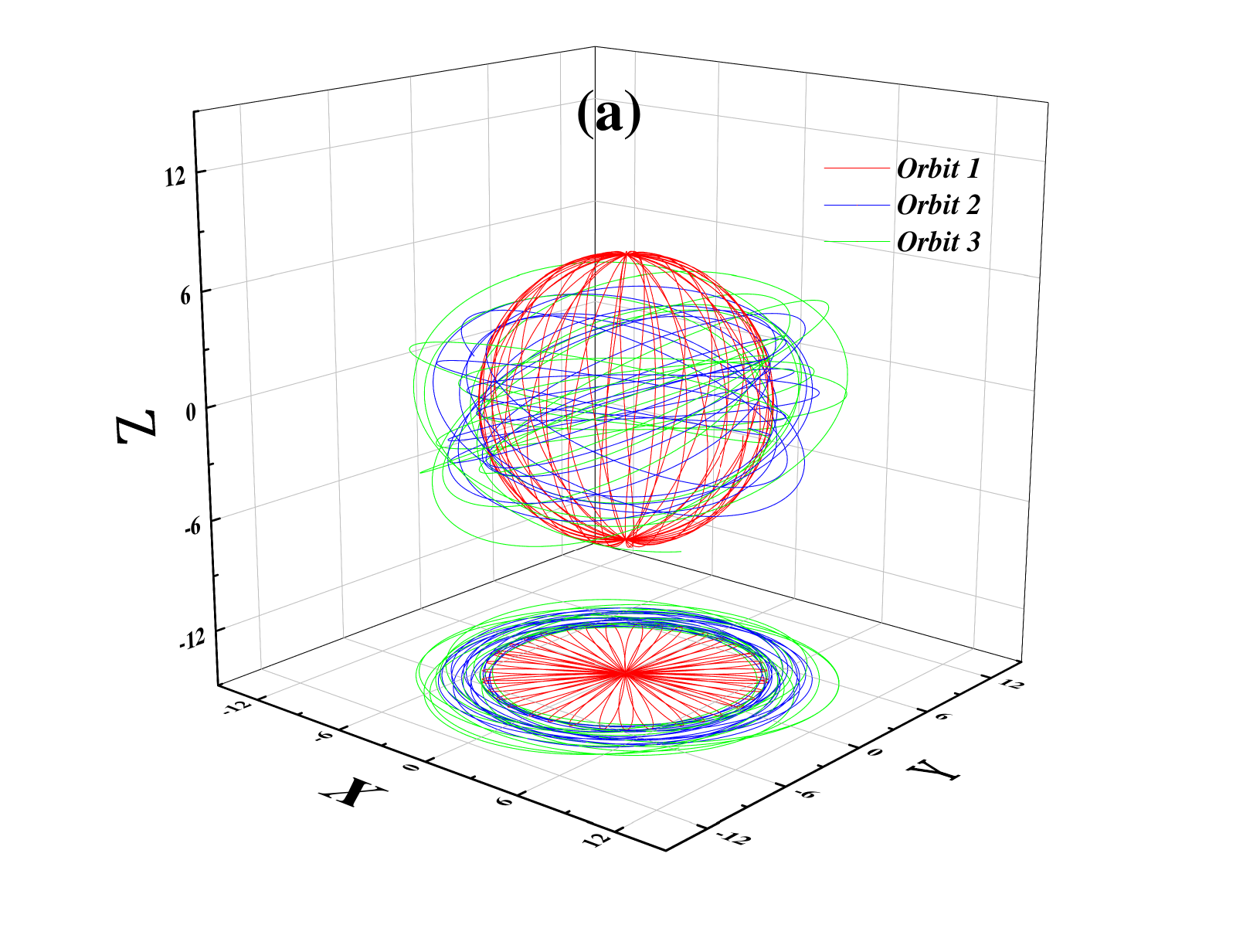}
\includegraphics[width=0.43  \textwidth]{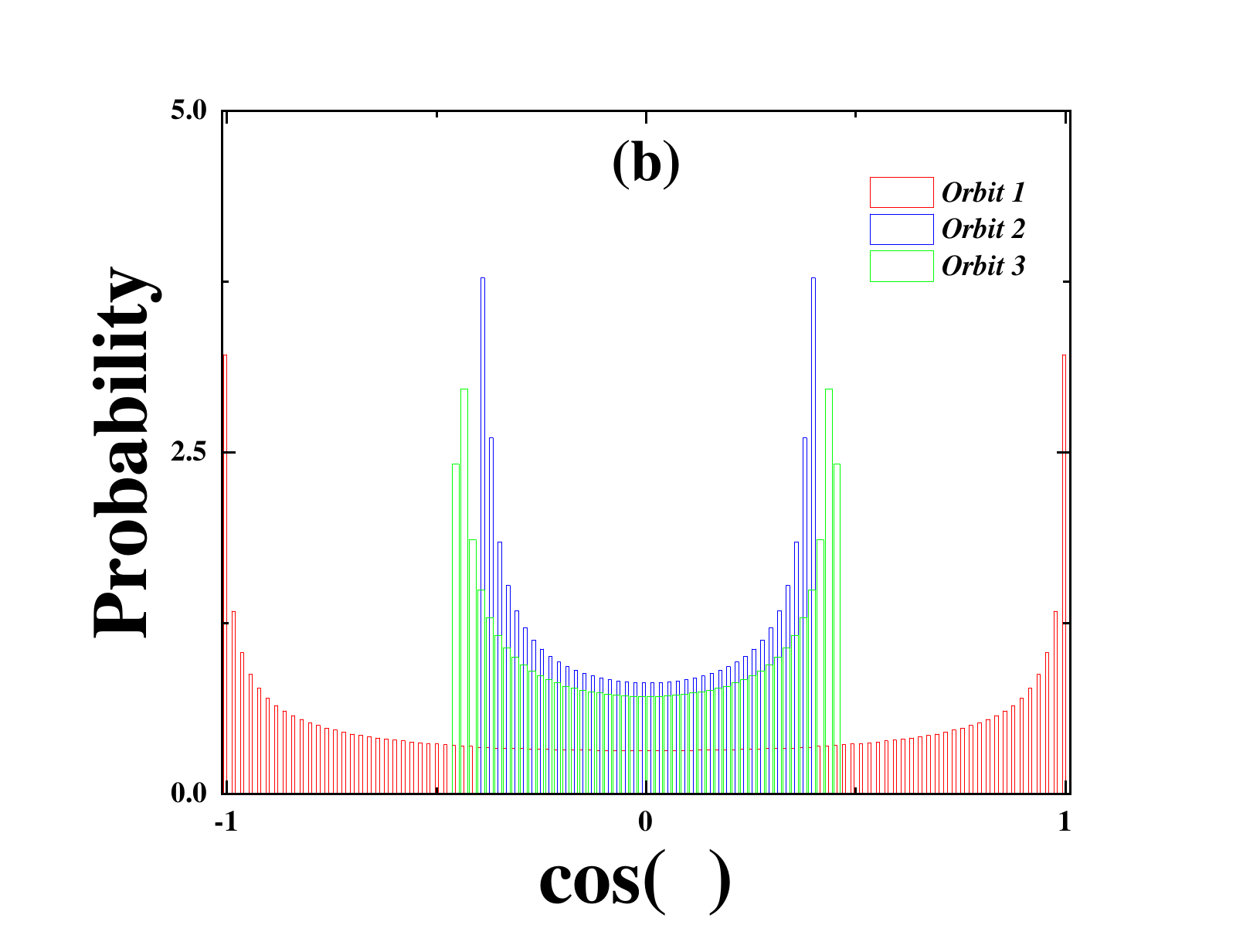}
\includegraphics[width=0.43  \textwidth]{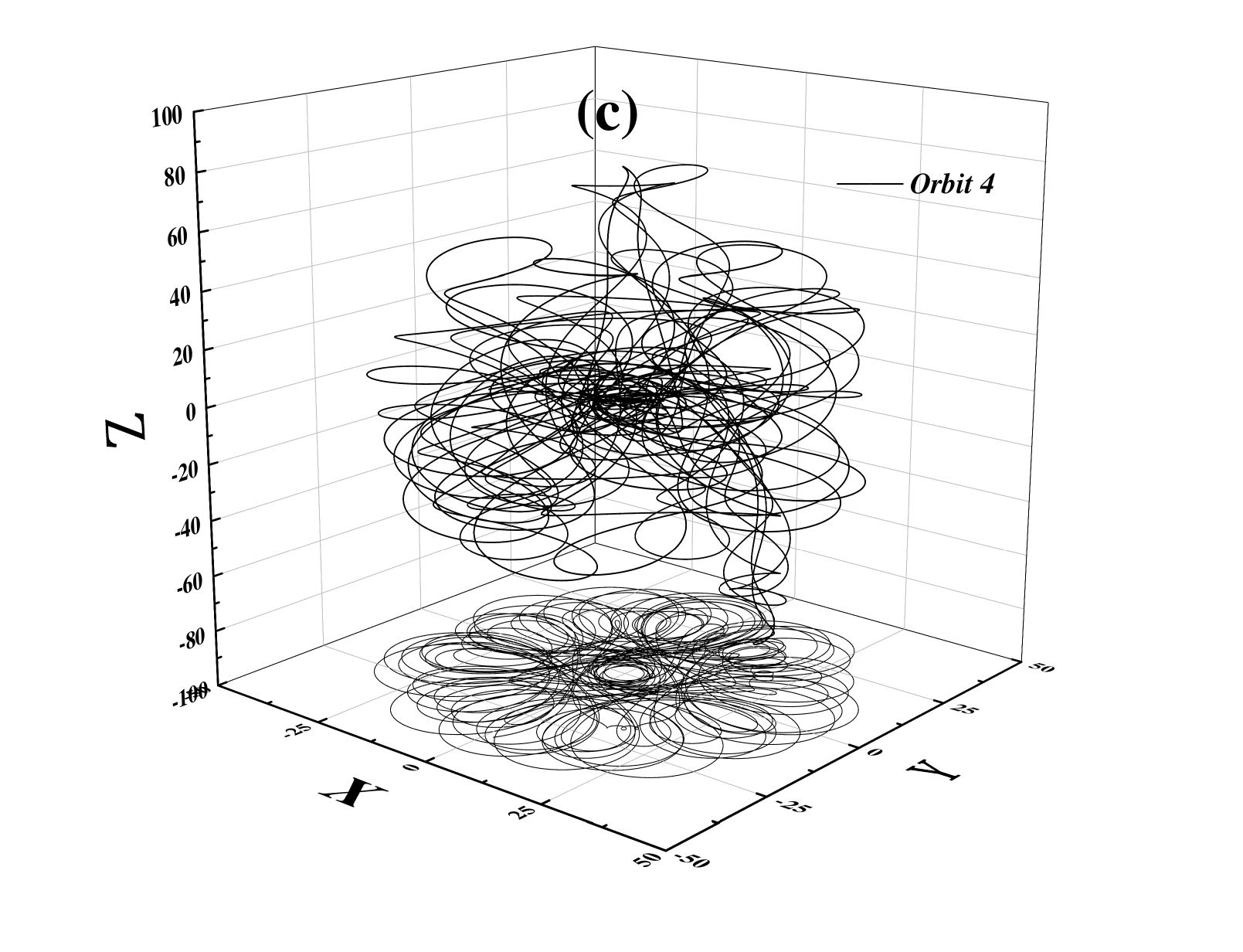}
\includegraphics[width=0.43  \textwidth]{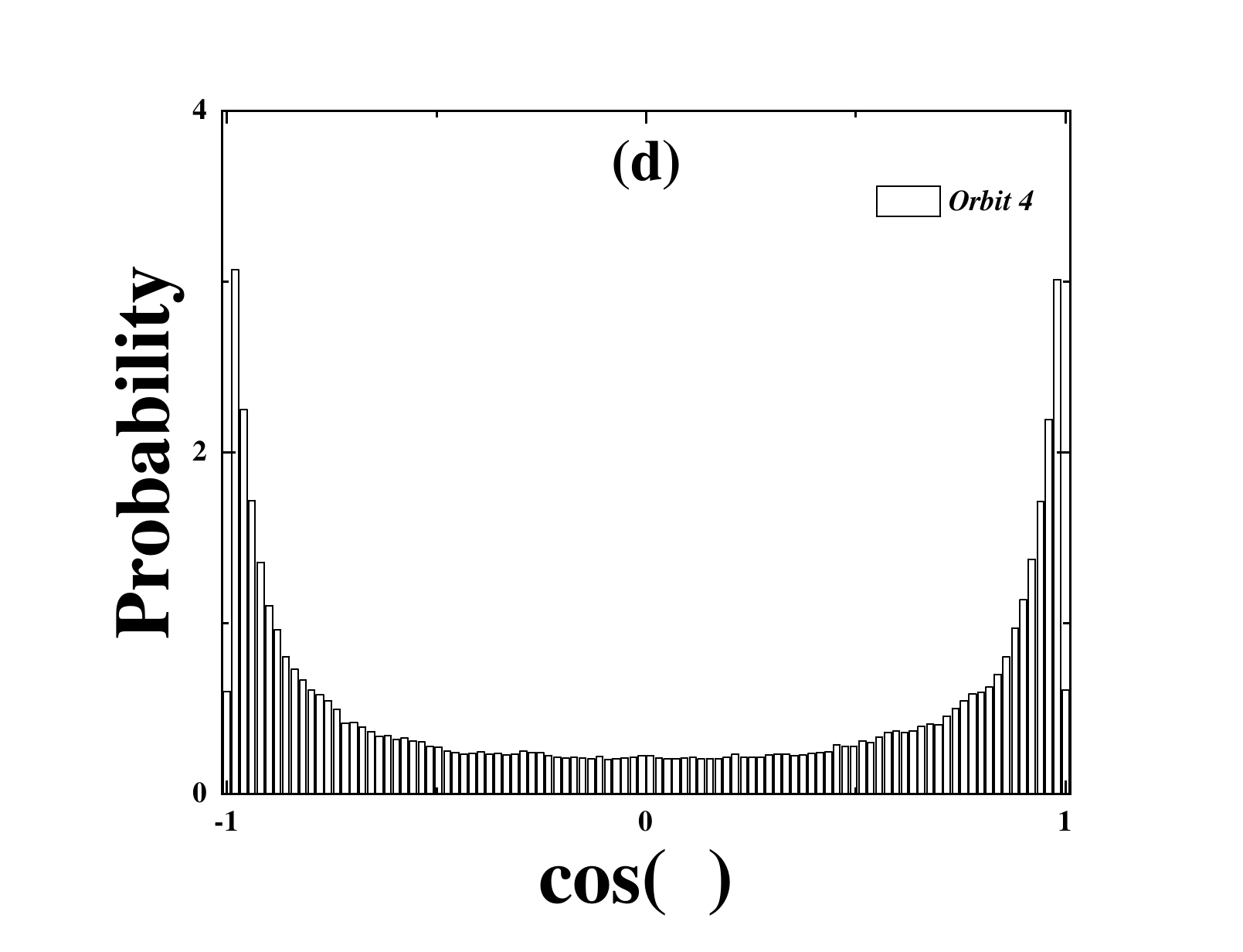}
\caption{Orbital trajectories and probability distributions in Kerr spacetime.
(a) and (b) The system parameters for Orbit 1 are $E=0.946$, $a=0.9$, $L_{z}=0$, $r=7.535$, and $\theta=90^{\circ}$.
The system parameters for Orbit 2 are $E=0.946$, $a=0.9$, $L_{z}=3$, $r=7.535$, and $\theta=90^{\circ}$.
The system parameters for Orbit 3 are $E=0.95$, $a=0.9$, $L_{z}=3$, $r=7.535$, and $\theta=90^{\circ}$.
(c) and (d) The system parameters for Orbit 4 are $E=0.998$, $a=0.9$, $L_{z}=3$, B=0.01, $r=7.535$, and $\theta=90^{\circ}$.}
 \label{Fig1}}
\end{figure*}

We assume that in the Kerr spacetime background, the PDF of the particle in the polar angle direction remains Eq. (12).
Using the effective potential method \cite{Cao:2022bvu}, we obtain three types of orbits: the polar orbit (Orbit 1) and two quasi-periodic orbits (Orbit 2 and Orbit 3).
As shown in Figs. 3(a) and (b), the order orbits exhibit a regular structure in three-dimensional space, with the polar angle PDF displaying a double-peak structure symmetric about the equatorial
plane.
This demonstrates the system's dynamical integrability and symmetry under undisturbed conditions, which is in excellent agreement with the theoretical results.
However, when an external magnetic field perturbation is introduced, although the corresponding PDF still maintains a macroscopic approximate symmetry, significant fluctuations appear on the local
scale (see Fig. 3(d)).

\begin{figure*}[htbp]
\center{
\includegraphics[width=0.3  \textwidth]{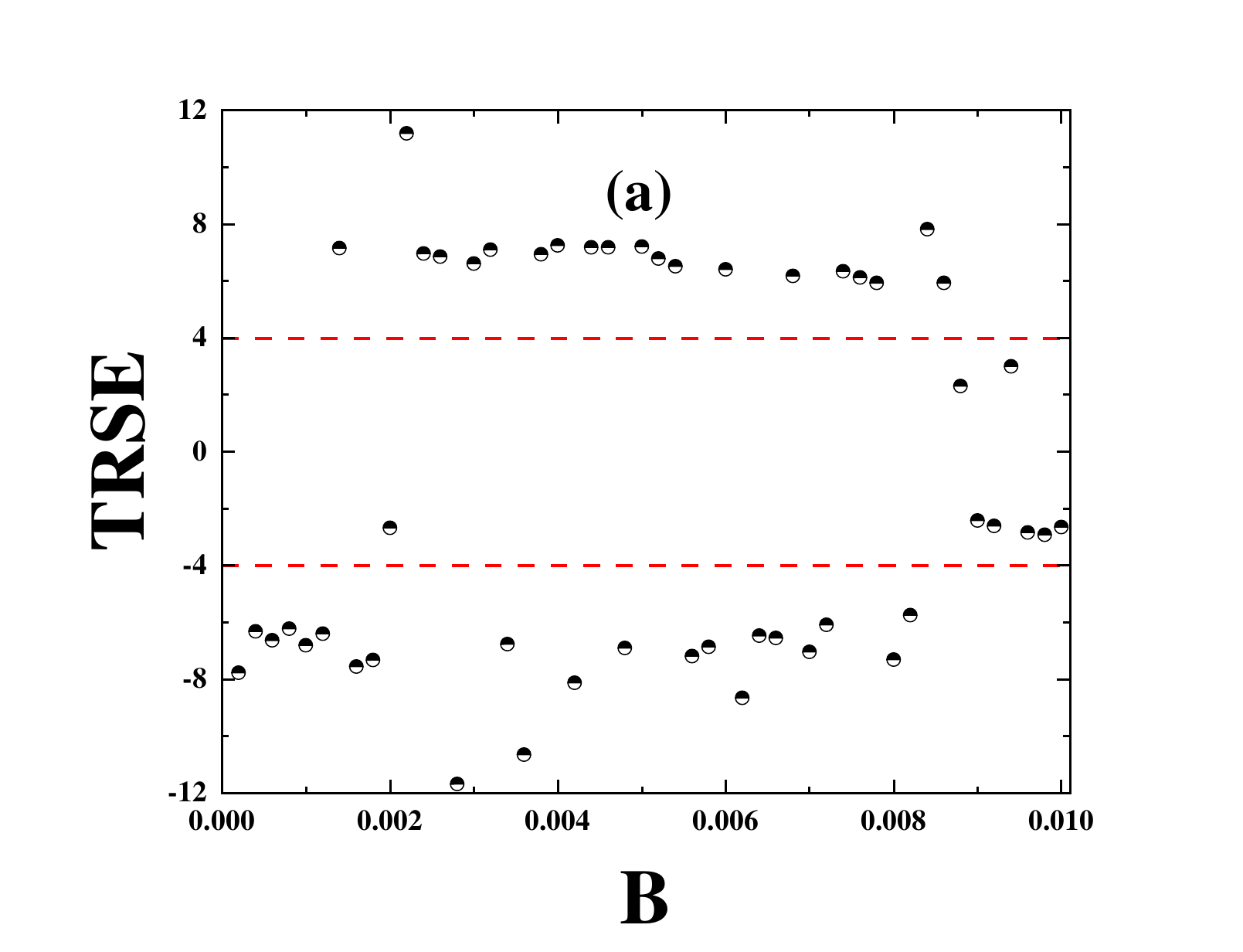}
\includegraphics[width=0.3  \textwidth]{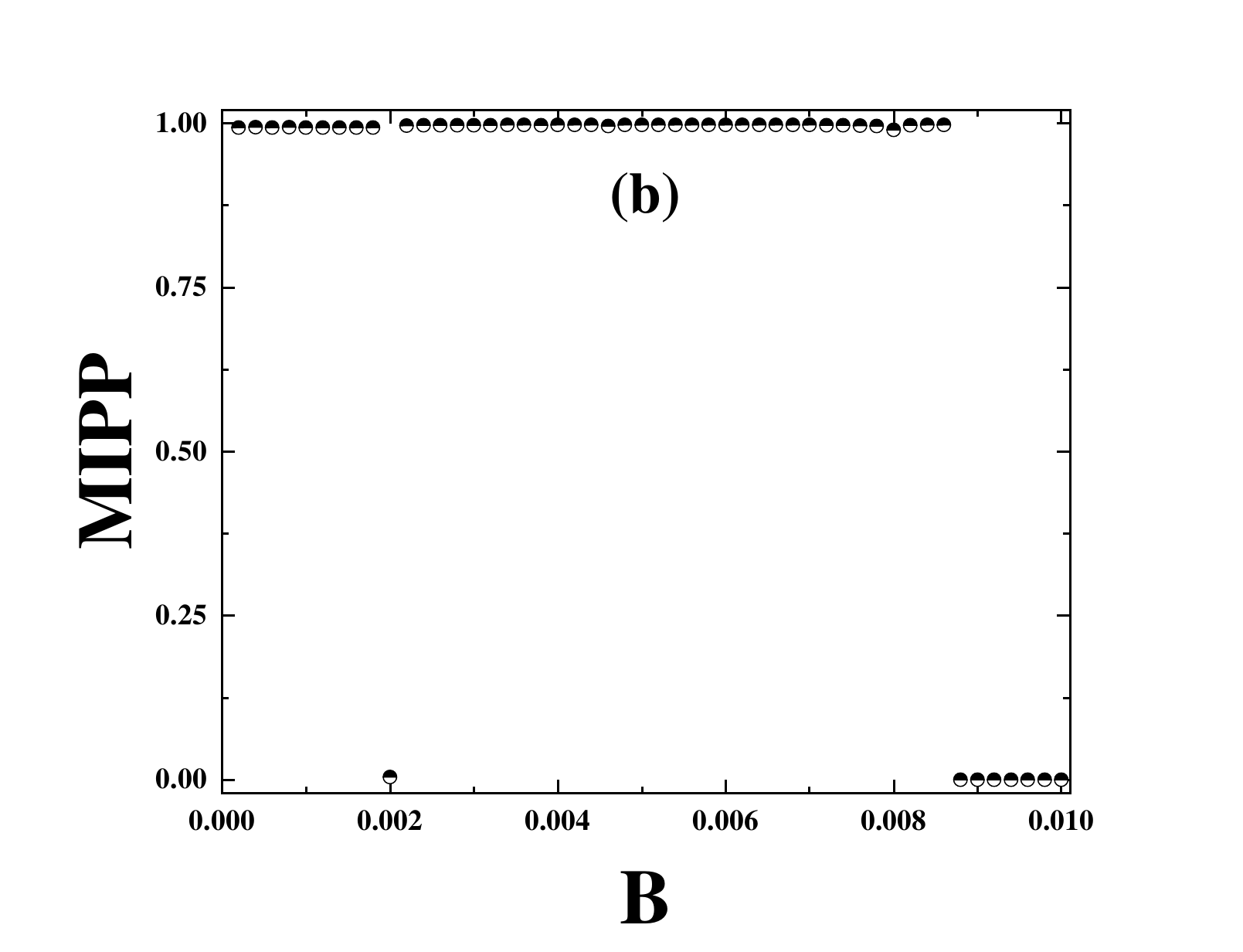}
\includegraphics[width=0.3  \textwidth]{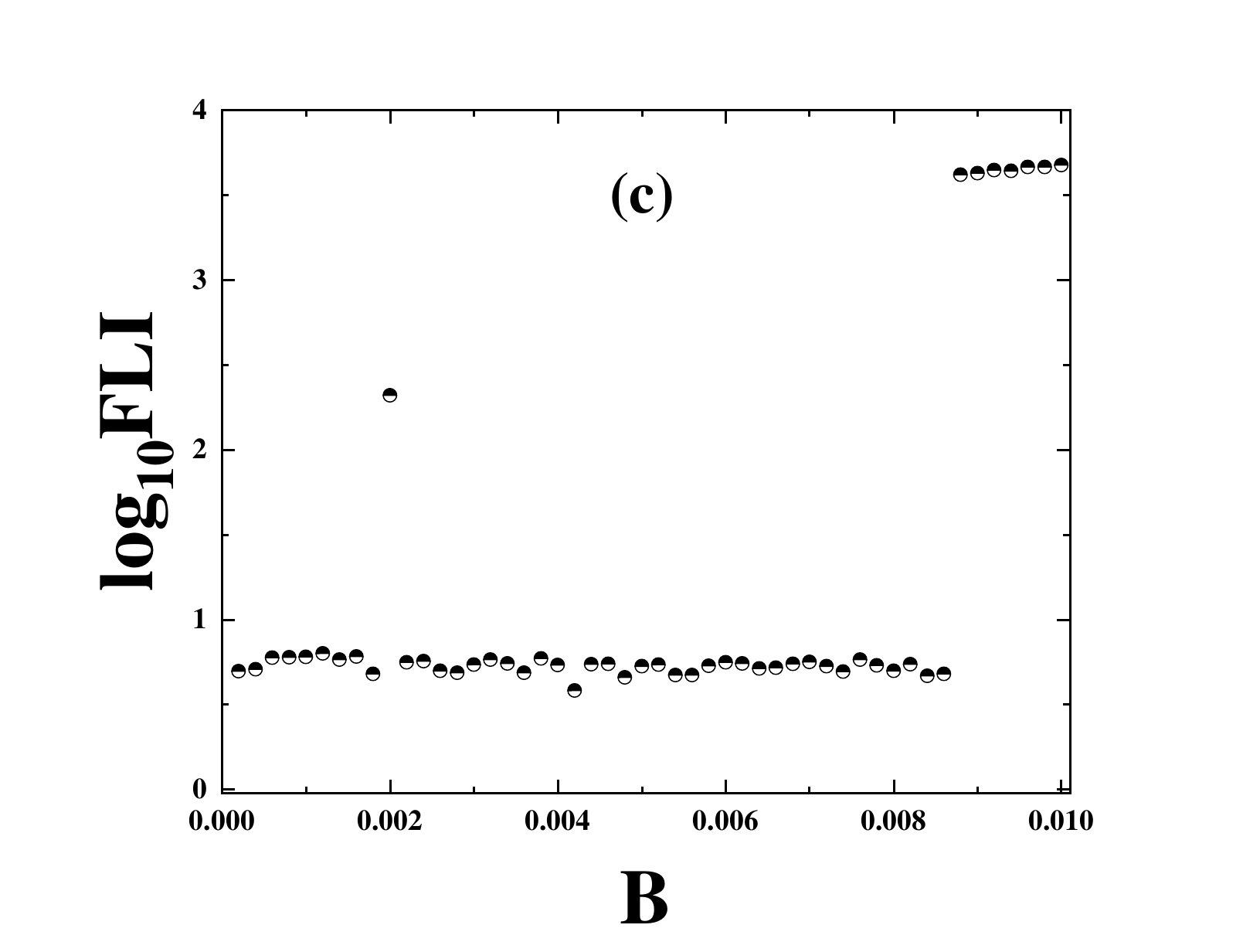}
\includegraphics[width=0.3  \textwidth]{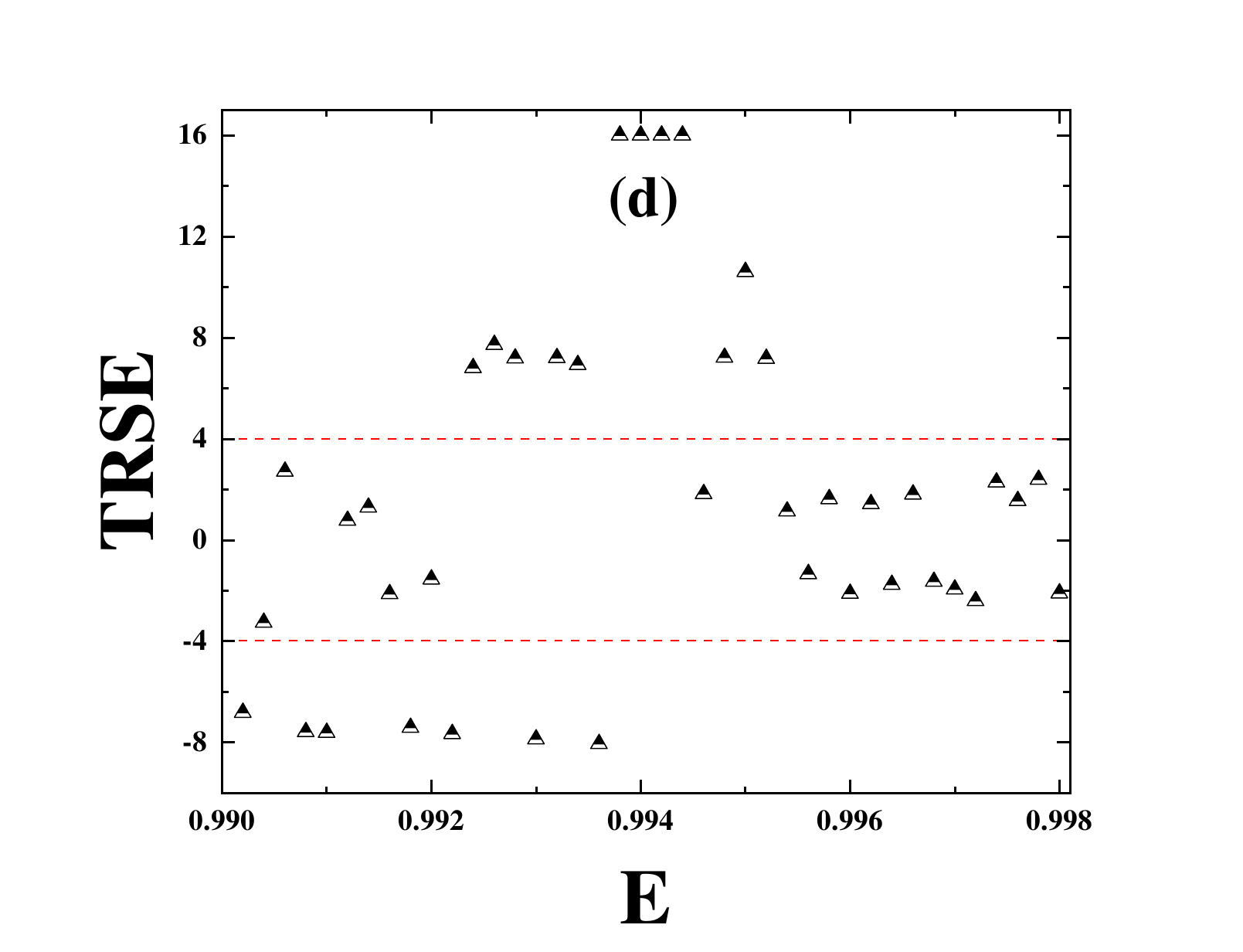}
\includegraphics[width=0.3  \textwidth]{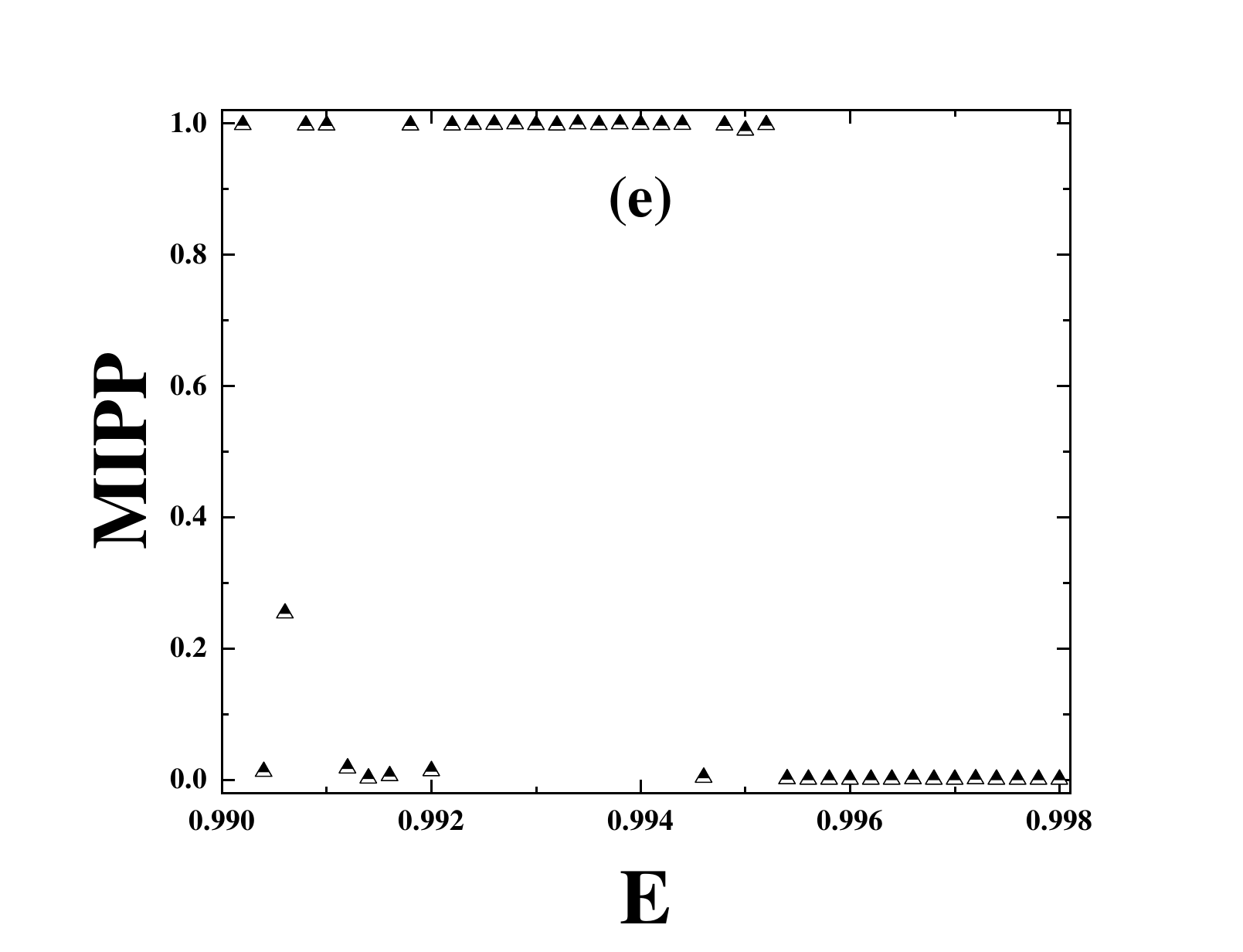}
\includegraphics[width=0.3  \textwidth]{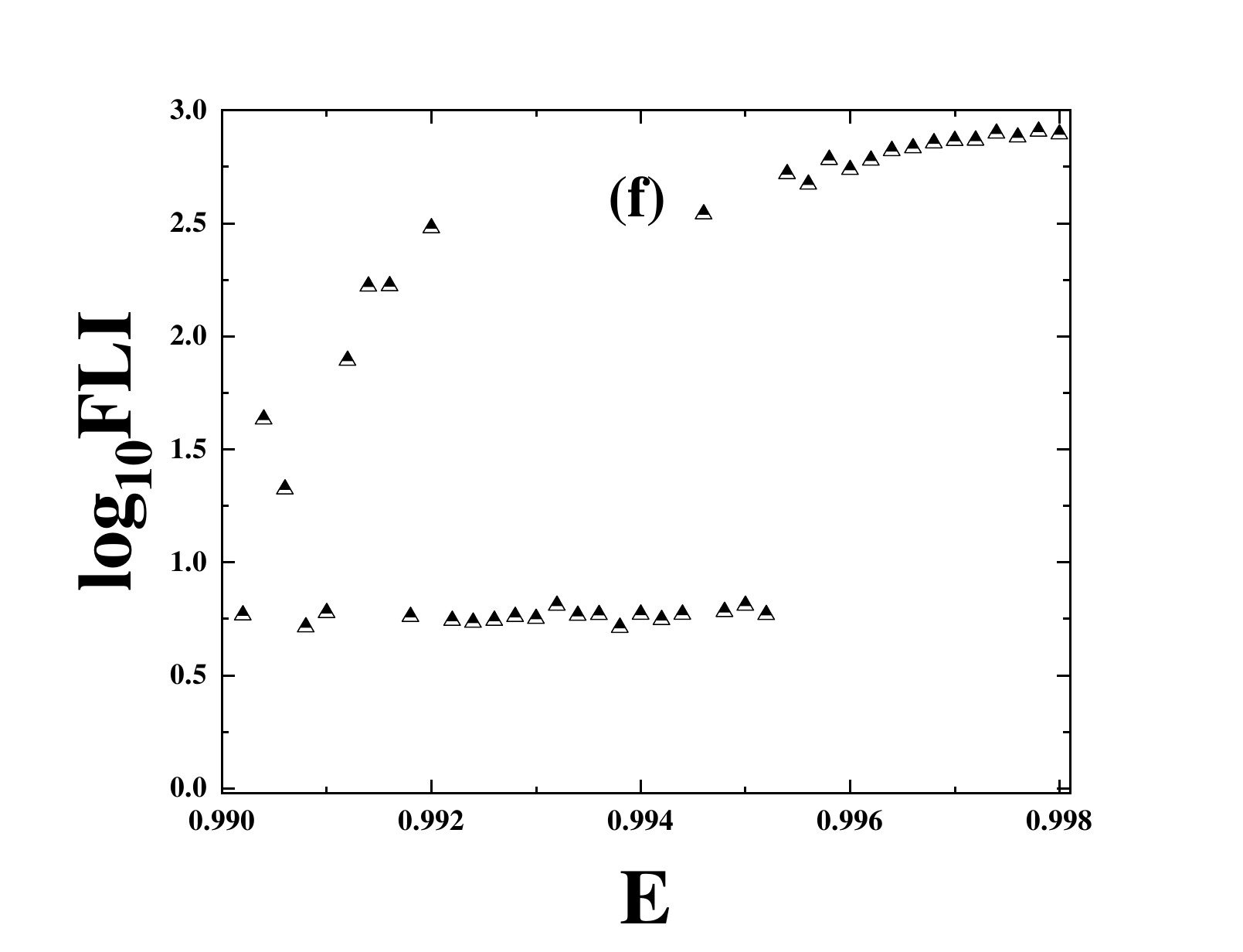}
\includegraphics[width=0.3  \textwidth]{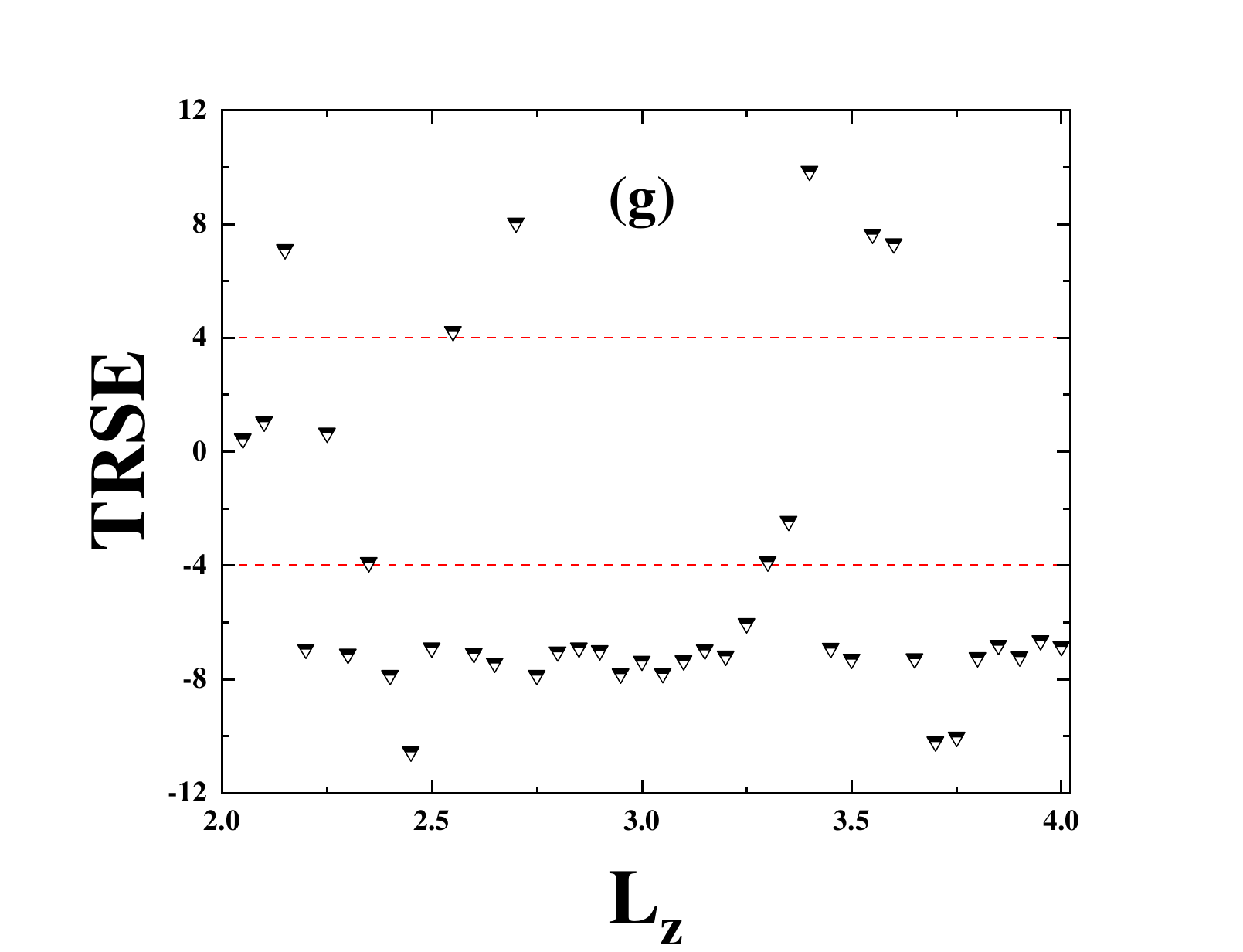}
\includegraphics[width=0.3  \textwidth]{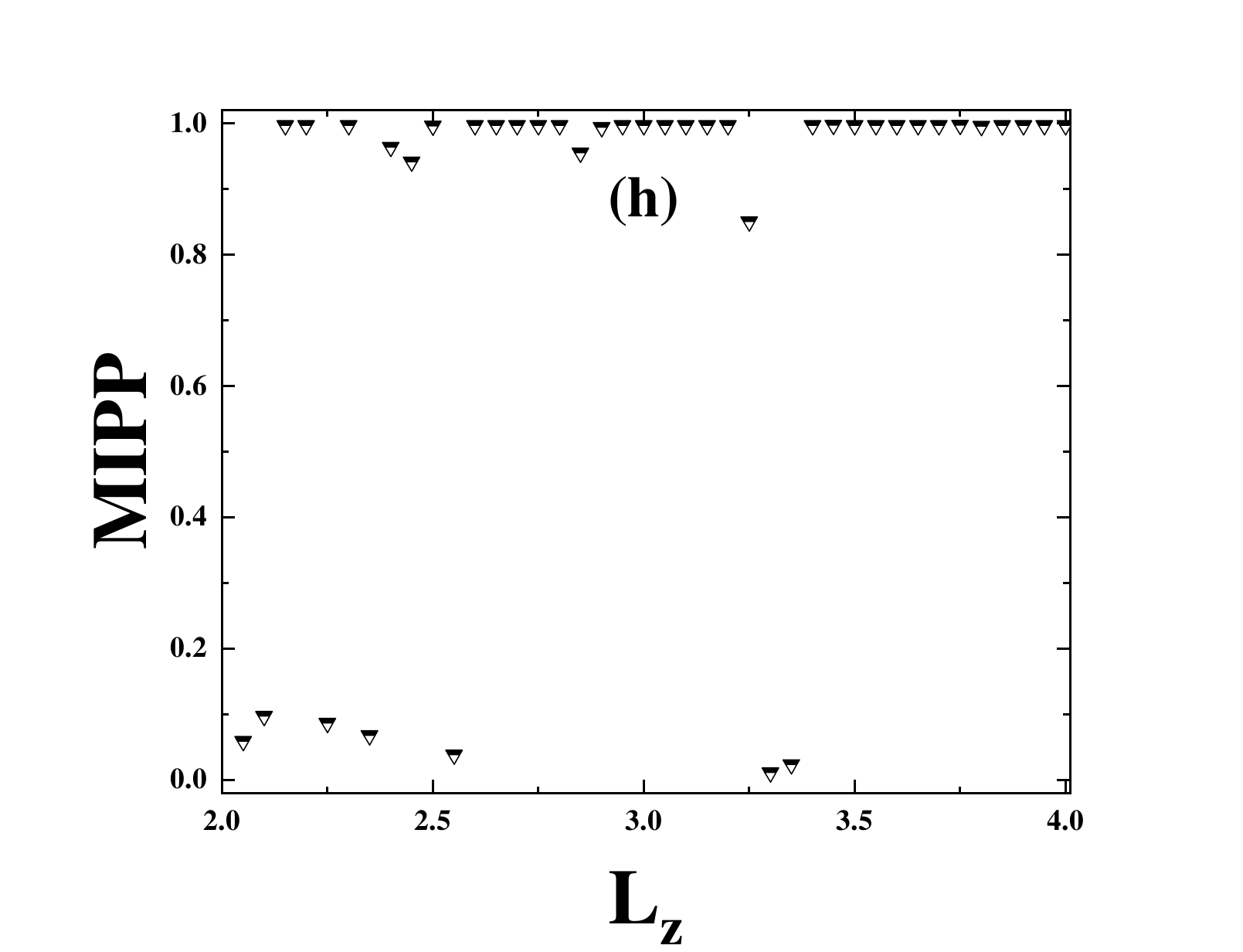}
\includegraphics[width=0.3  \textwidth]{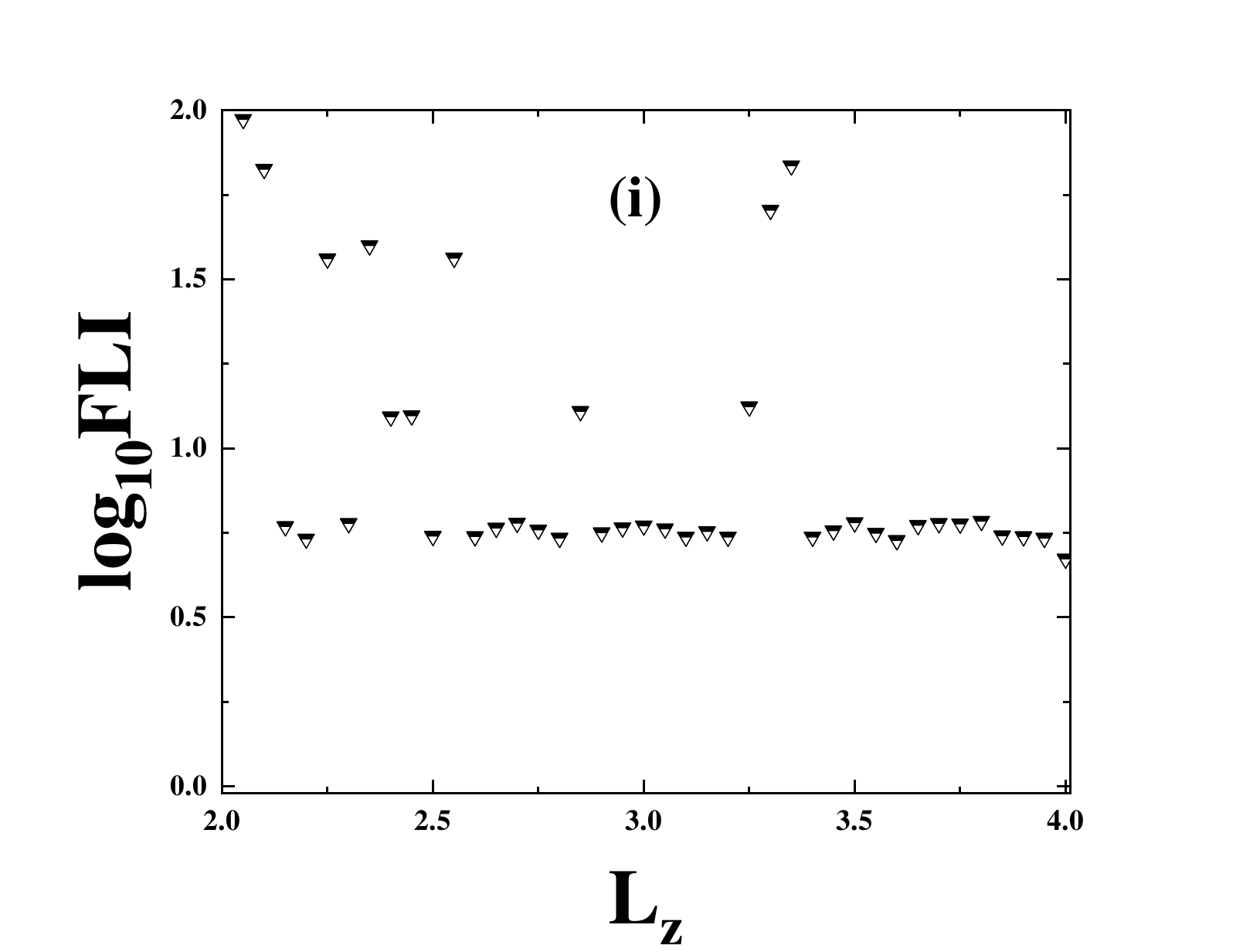}
\caption{TRSE, MIPP and FLI scanplots for the key parameter($B, E, L_{z}$) in Kerr spacetime. (a)-(c) The range of the magnetic field parameter B is from 0.0002 to 0.01,
and a total of 50 orbits were calculated. The system parameters are $E=0.989$, $L_{z}=3$, $a=0.998$, $r=8$, and $\theta=90^{\circ}$.
(d)-(f) The range of the  parameter $E$ is from 0.9902 to 0.998,
and a total of 40 orbits were calculated. The system parameters are $L_{z}=4$, $B=0.001$, $a=0.01$, $r=10$, and $\theta=90^{\circ}$.
(g)-(i) The range of the  parameter $L_{z}$ is from 2.05 to 4,
and a total of 40 orbits were calculated. The system parameters are $E=0.99$, $B=0.001$, $a=0.01$, $r=11$, and $\theta=90^{\circ}$.}
 \label{Fig1}}
\end{figure*}

Fig. 4 presents a comparative analysis of TRSE, MIPP and FLI in response to key system parameters (magnetic field parameter $B$, energy $E$, and angular momentum $L_{z}$ )\cite{Tancredi:2001,
Wu:2003pe, Wu:2006rx,Cardoso:2008bp,Ma:2014aha,Lei:2020clg}.
The scan results demonstrate the consistency among these chaos indicators in identifying chaotic orbit states.
The breakdown of time-reversal symmetry, as measured by the TRSE value approaching zero, coincides with the loss of statistical correlation in the orbit (indicated by the MIPP value tending to zero) across the entire parameter space.
This correspondence is especially evident in logarithmic coordinates: the TRSE values for chaotic orbits differ from those of order orbits by more than two orders of magnitude, with chaotic orbit
data points concentrated near the red dashed-line threshold, while order orbit data points are widely distributed in regions far from the threshold. Additionally, the sign of the TRSE value
provides
a more refined characterization of the orbit's complexity.
\subsection{Application to the Schwarzschild-Melvin Black Hole}

The Schwarzschild-Melvin solution is an exact solution to the Einstein-Maxwell equations, describing a Schwarzschild black hole immersed in a uniform magnetic field.
Compared to the Schwarzschild spacetime, this model is non-integrable, resulting in chaotic behavior in the orbits of photons \cite{Kostaros:2021usv},
\begin{eqnarray}
ds^{2}=g_{ab}dx^{a}dx^{b}=\Lambda^{2}[\frac{2M-r}{r}dt^{2}+\frac{rdr^{2}}{r-2M}+r^{2}d\theta^{2}]+\frac{r^{2}}{\Lambda^{2}}\sin^{2}\theta d\phi^{2},
\end{eqnarray}
where $\Lambda=1+\frac{1}{4}B^{2}r^{2}\sin^{2}\theta$, $B$ is the magnetic field parameter. As shown in Fig. 5, we scan the particle energy parameter $E$ in the non-integrable Schwarzschild-Melvin
black hole spacetime, and compare the performance of Shannon entropy, TRSE, and MIPP in identifying the chaotic characteristics of photon orbits.
Fig. 5(a) implies the limitations of traditional Shannon entropy as a chaos criterion. Although the $H(X)$ values for chaotic orbits (red data points) are generally higher,
there is a clear counterexample near
$E\approx0.61$, where the entropy value for chaotic orbits is lower than that for order orbits (blue data points).
This phenomenon suggests that chaotic motion does not always correspond to higher statistical uncertainty, and the numerical value of Shannon entropy alone cannot reliably distinguish between
dynamic states.
The scan results in Figs. 5(b) and (c) demonstrate the high consistency and effectiveness of the TRSE and MIPP indicators.
Together, they capture the essential characteristics of chaotic motion from the perspectives of information theory and symmetry.
Fig. 5(d) provides an intuitive physical explanation for the above chaos indicators through the PDFs.
For order orbits, the PDF exhibits a perfect symmetric distribution along the equatorial plane.
In contrast, while the PDF for chaotic orbits appears approximately symmetric on a macroscopic scale, significant fluctuations occur on a local scale.
More significantly, the distribution deviates notably from theoretical predictions based on integrable spacetimes.
This strongly indicates that in non-integrable spacetimes, like the Schwarzschild-Melvin spacetime, the statistical behavior of particle orbits exhibits a richer diversity.

\begin{figure*}[htbp]
\center{
\includegraphics[width=0.43  \textwidth]{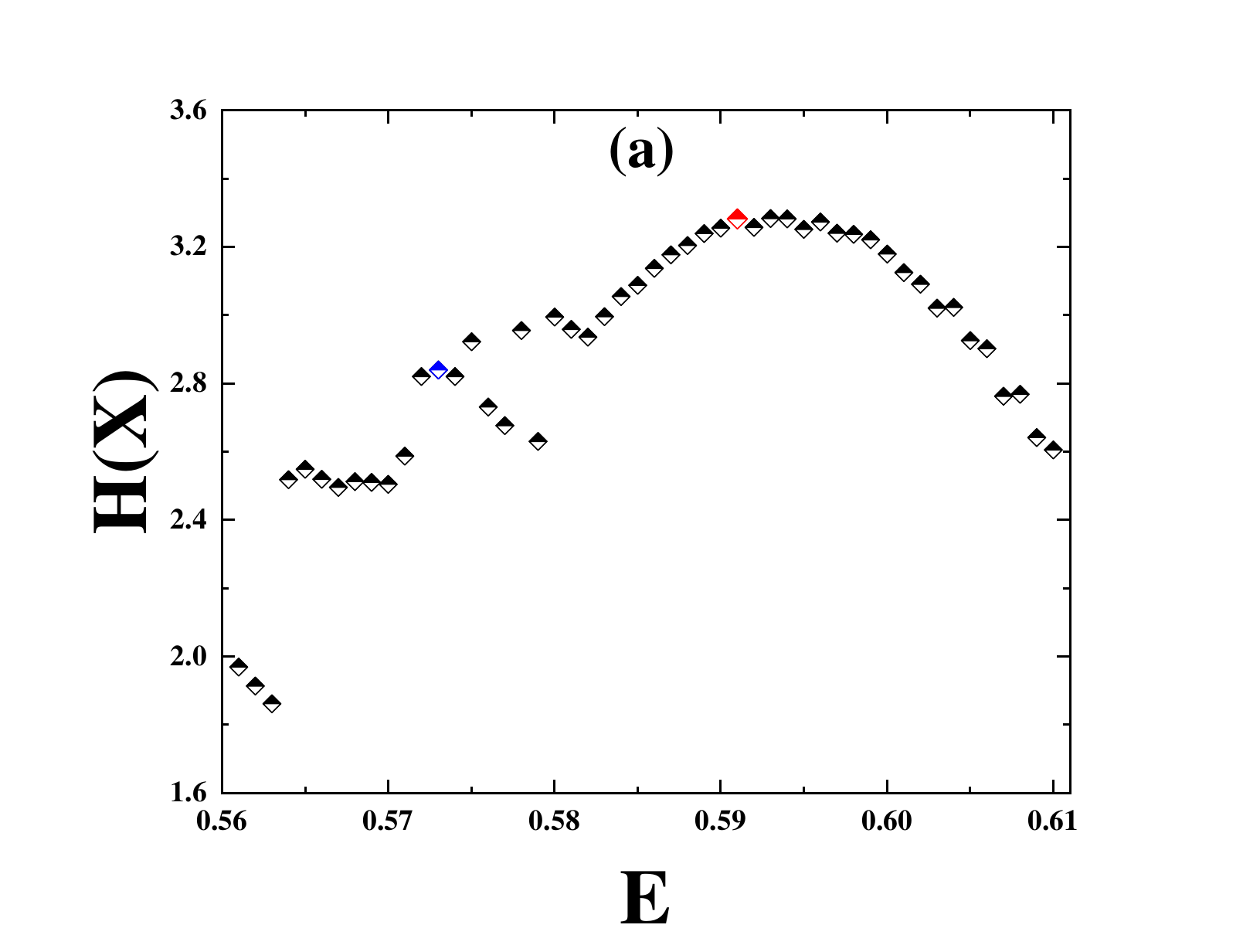}
\includegraphics[width=0.43  \textwidth]{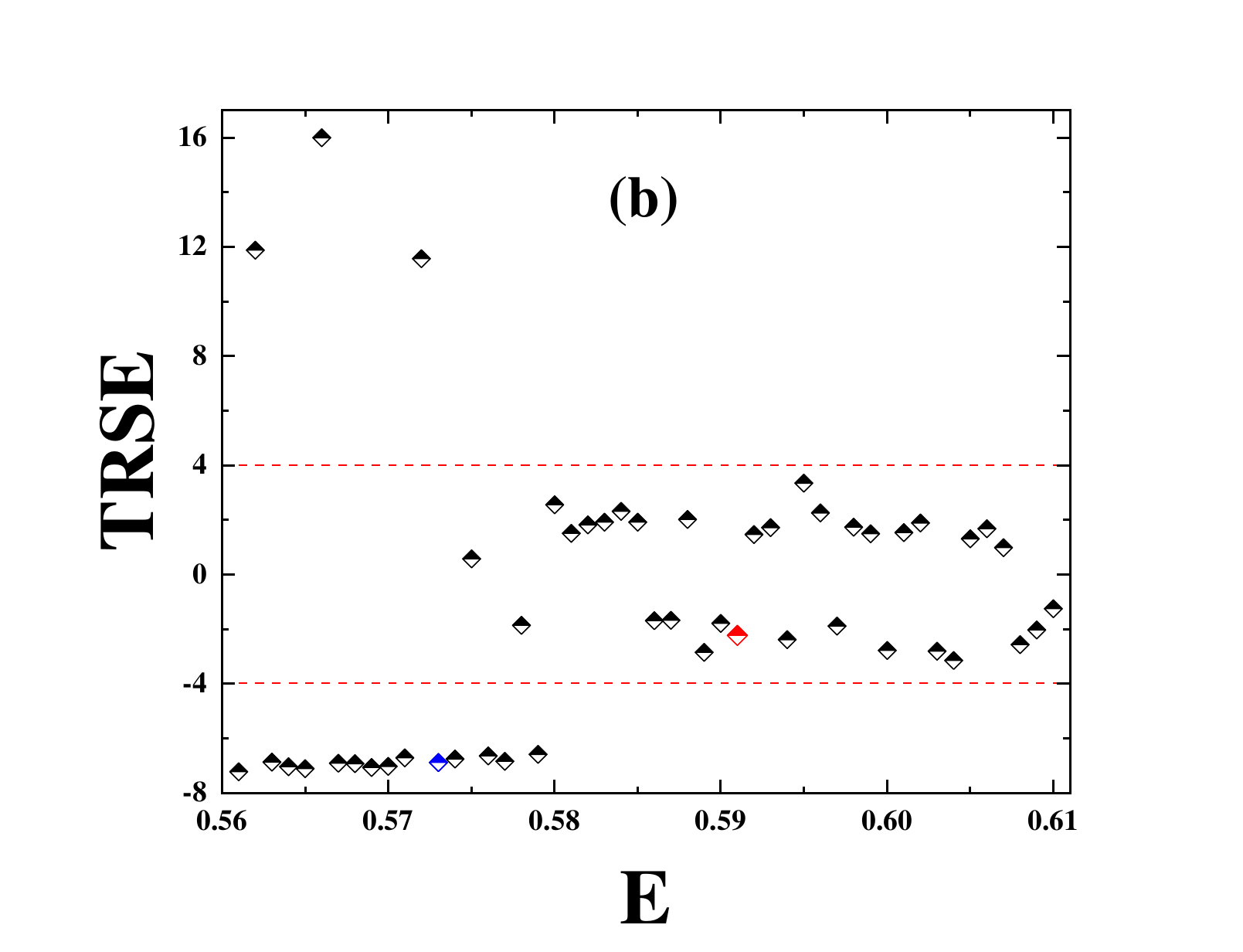}
\includegraphics[width=0.43  \textwidth]{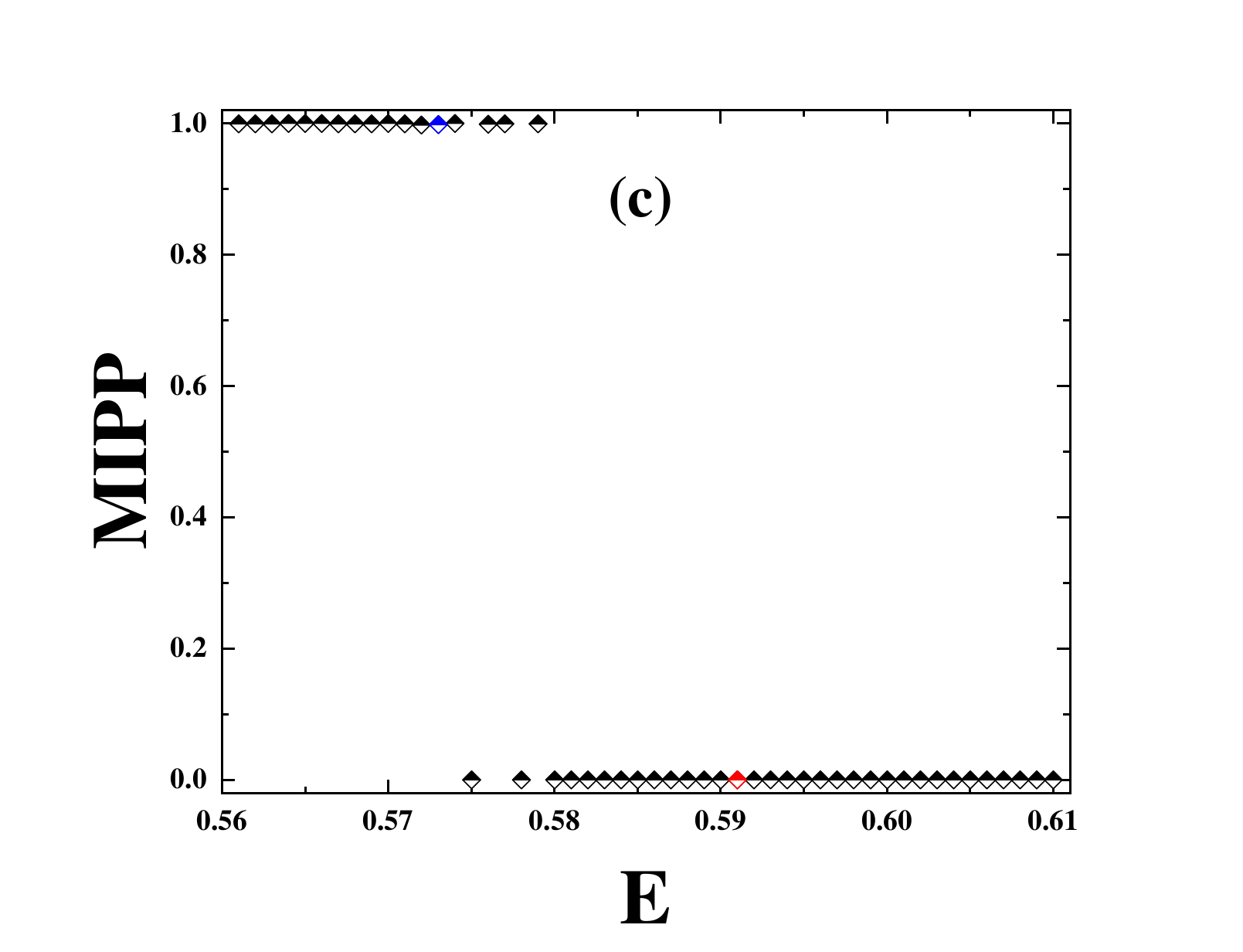}
\includegraphics[width=0.43  \textwidth]{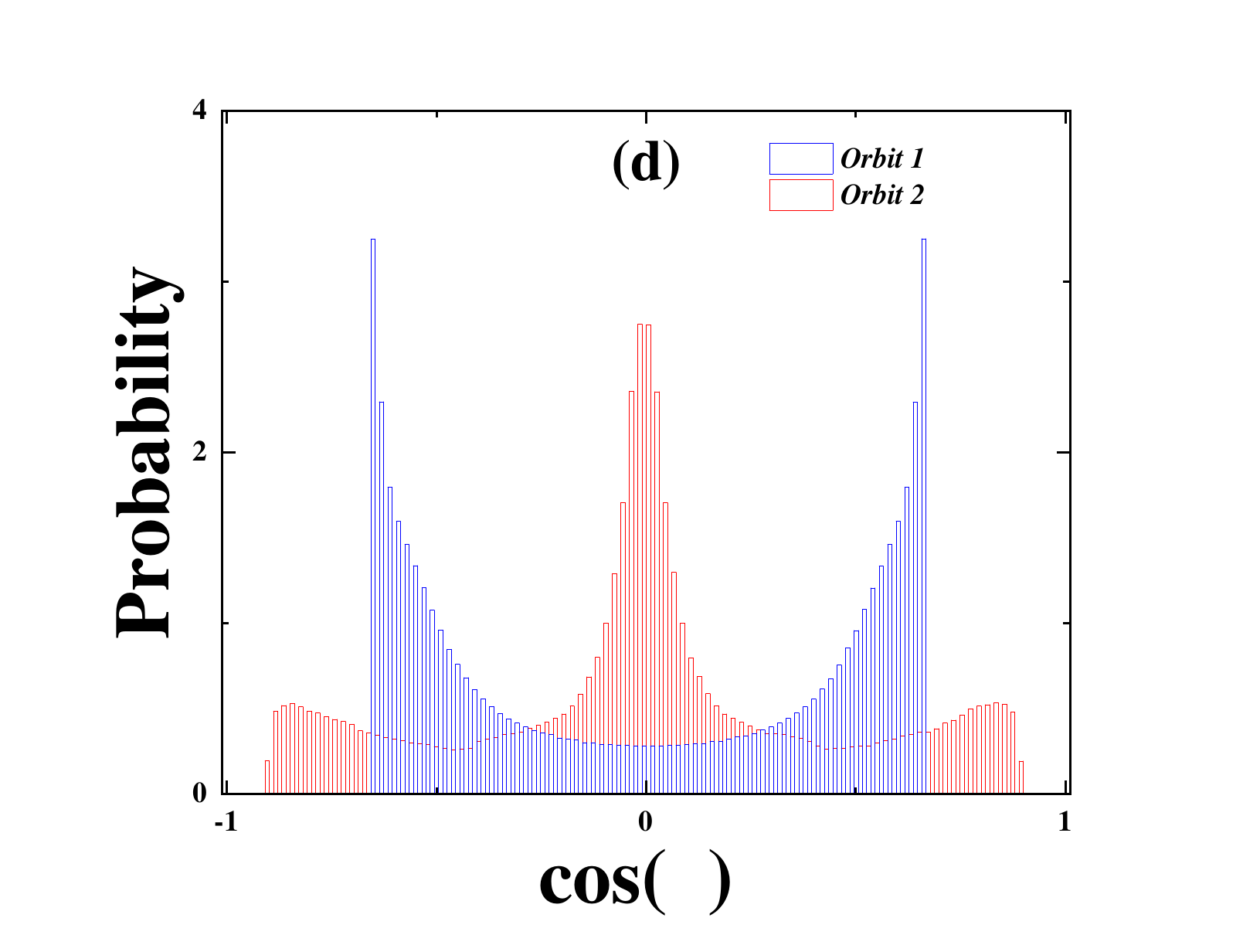}
\caption{MIPP, TRSE, and Shannon entropy scanplots for the parameter $E$ and orbital probability distributions in Schwarzschild-Melvin spacetime.
The range of the  parameter $E$ is from 0.561 to 0.61, and a total of 50 orbits were calculated. The system parameters are $B=0.1$, $L_{z}=4$, $r=10.656$, and $\theta=90^{\circ}$.}
 \label{Fig5}}
\end{figure*}

\section{Conclusions}\label{sec:four}
In this work, we introduce the time-reversed Shannon entropy (TRSE) as a novel and efficiency indicator for detecting chaotic dynamics in non-integrable general relativistic systems. The principal innovation of TRSE lies in its integration of the fundamental concept of time-reversal symmetry breaking with information-theoretic entropy, enabling a robust distinction between regular and chaotic orbital evolution through quantitative comparison of forward and backward temporal dynamics. Our numerical validations, conducted in both Kerr and Schwarzschild-Melvin spacetimes, confirm that TRSE reliably identifies chaos by capturing significant statistical discrepancies under time reversal, whereas traditional Shannon entropy alone may lead to ambiguities in interpretation.

Concurrently, we have refined the mutual information for a particle pair (MIPP) method, enhancing its applicability as a complementary chaos probe. Scans across broad parameter regions reveal a strong quantitative agreement between TRSE and MIPP, underscoring the consistency and reliability of both indicators. While TRSE quantifies symmetry breaking in single-orbit evolution, MIPP measures statistical correlations between neighboring trajectories. Together, they establish a unified and efficient framework for chaos diagnosis in curved spacetimes.

Looking ahead, the TRSE and MIPP methodologies hold promise for broader applications. These include extending their use to black hole solutions in modified gravity theories, providing new criteria for testing theoretical models, as well as applying them to gravitational waveform simulations for extreme mass-ratio inspirals, where chaotic motion may imprint observable effects. Furthermore, a quantum generalization of these tools could open pathways to probing quantum chaos in non=integrable systems. Such developments would significantly advance our understanding of chaotic dynamics from classical to quantum regimes.

\section*{Acknowledgments}
This work is supported by the National Natural Science
Foundation of China (NSFC) under Grant nos. 12235019 and 12275106.


\begin{thebibliography}{99}

\bibitem{Clausius:1865}
R. Clausius,
``Ueber verschiedene f\"{u}r die Anwendung bequeme Formen der Hauptgleichungen der mechanischen W\"{a}rmetheorie,''
Ann. Phys.  \textbf{201} (1865) 353-400.
doi:10.1002/andp.18652010702.


\bibitem{Boltzmann1897}
Ludwig Boltzmann,
``Vorlesungen \"{u}ber Gastheorie. Erster Theil: Theorie der Gase mit einatomigen Molek\"{u}len, deren Dimensionen gegen die mittlere Wegl\"{a}nge verschwinden,''
J. A. Barth, Leipzig (1897).

\bibitem{Shannon1948}
Claude E. Shannon,
``A Mathematical Theory of Communication,''
Bell System Technical Journal \textbf{27} (1948).
doi:10.1002/j.1538-7305.1948.tb01338.x.

\bibitem{VonNeumann1932}
John von Neumann,
``Mathematische Grundlagen der Quantenmechanik,''
Springer-Verlag, Berlin (1932).
doi:10.1007/978-3-642-61409-5.



\bibitem{Bekenstein:1973ur}
J.~D.~Bekenstein,
``Black holes and entropy,''
Phys. Rev. D \textbf{7}, 2333-2346 (1973)
doi:10.1103/PhysRevD.7.2333



\bibitem{Hawking:1974rv}
S.~W.~Hawking,
`Black hole explosions,''
Nature \textbf{248}, 30-31 (1974)
doi:10.1038/248030a0


\bibitem{LIGOScientific:2025rid}
A.~G.~Abac \textit{et al.} [LIGO Scientific, Virgo and KAGRA],
``GW250114: Testing Hawking{\textquoteright}s Area Law and the Kerr Nature of Black Holes,
Phys. Rev. Lett. \textbf{135}, no.11, 111403 (2025)
doi:10.1103/kw5g-d732
[arXiv:2509.08054 [gr-qc]].



\bibitem{Seifert:2005rlb}
U.~Seifert,
``Entropy Production along a Stochastic Trajectory and an Integral Fluctuation Theorem,''
Phys. Rev. Lett. \textbf{95}, no.4, 040602 (2005)
doi:10.1103/PhysRevLett.95.040602


\bibitem{Leggio:2013wrr}
B.~Leggio, A.~Napoli, A.~Messina and H.~P.~Breuer,
``Entropy production and information fluctuations along quantum trajectories,''
Phys. Rev. A \textbf{88}, 042111 (2013)
doi:10.1103/PhysRevA.88.042111
[arXiv:1305.6733 [quant-ph]].


\bibitem{Cao:2024bjk}
W.~Cao, Y.~Huang and H.~Zhang,
``Screen chaotic motion by Shannon entropy in curved spacetimes,''
Eur. Phys. J. C \textbf{85}, no.5, 568 (2025)
doi:10.1140/epjc/s10052-025-14310-x
[arXiv:2410.20870 [gr-qc]].

\bibitem{Cao:2024rvo}
W.~Cao, Y.~Huang and H.~Zhang,
``Mutual information for a particle pair and its application to diagnose chaos in curved spacetime,''
Phys. Rev. E \textbf{111}, no.5, 054207 (2025)
doi:10.1103/PhysRevE.111.054207
[arXiv:2412.16931 [gr-qc]].




\bibitem{Wald:1974np}
R.~M.~Wald,
``Black hole in a uniform magnetic field,''
Phys. Rev. D \textbf{10}, 1680-1685 (1974)
doi:10.1103/PhysRevD.10.1680


\bibitem{Karas:1992}
Karas, V., Vokrouhlick\'{y}, D.,
``Chaotic motion of test particles in the Ernst space-time,''
%Gen Relat Gravit \textbf{24}, 729743 (1992),
https://doi.org/10.1007/BF00760079.



\bibitem{Li:2018wtz}
D.~Li and X.~Wu,
``Chaotic motion of neutral and charged particles in a magnetized Ernst-Schwarzschild spacetime,''
Eur. Phys. J. Plus \textbf{134}, no.3, 96 (2019)
doi:10.1140/epjp/i2019-12502-9
[arXiv:1803.02119 [gr-qc]].


\bibitem{Cao:2024ihv}
W.~Cao, X.~Wu and J.~Lyu,
``Electromagnetic field and chaotic charged-particle motion around hairy black holes in Horndeski gravity,''
Eur. Phys. J. C \textbf{84}, no.4, 435 (2024)
doi:10.1140/epjc/s10052-024-12804-8
[arXiv:2404.19225 [gr-qc]].


\bibitem{Xu:2024ble}
 Z.~M.~Xu, D.~Z.~Ma,  W.~F.~Cao and K.~Li,
``Chaotic motion of charged test particles in a Kerr-MOG black hole with explicit symplectic algorithms,''
Eur. Phys. J. C \textbf{85}, no.7, 770 (2025)
doi:10.1140/epjc/s10052-025-14425-1
[arXiv:2412.06122 [gr-qc]].


\bibitem{Li:2025jfq}
D.~Li, Y.~Zuo, S.~Hu, C.~Deng, Y.~Wang and W.~Cao,
``Shadows of three black holes in static equilibrium configuration,''
Eur. Phys. J. C \textbf{85}, no.8, 905 (2025)
doi:10.1140/epjc/s10052-025-14654-4
[arXiv:2504.04102 [gr-qc]].




\bibitem{Benenti:1979erw}
S.~Benenti and M.~Francaviglia,
``Remarks on certain separability structures and their applications to general relativity,''
Gen. Rel. Grav. \textbf{10}, no.1, 79-92 (1979)
doi:10.1007/bf00757025

\bibitem{Papadopoulos:2018nvd}
G.~O.~Papadopoulos and K.~D.~Kokkotas,
``Preserving Kerr symmetries in deformed spacetimes,''
Class. Quant. Grav. \textbf{35}, no.18, 185014 (2018)
doi:10.1088/1361-6382/aad7f4
[arXiv:1807.08594 [gr-qc]].


\bibitem{Masoliver:2011}
J.~Masoliver and A.~Ros,
``Integrability and chaos: the classical uncertainty,''
Eur. J. Phys. \textbf{32}, no.2, 431-458 (2011)
doi:10.1088/0143-0807/32/2/016
[arXiv:1012.4384 [nlin.CD]].



\bibitem{1972PhRvD...5..814W}
D.~C.~Wilkins,
``Bound Geodesics in the Kerr Metric,''
Phys. Rev. D \textbf{5}, no.~4, 814-822 (1972)
doi:10.1103/PhysRevD.5.814.


\bibitem{1972ApJ...178..347B}
J.~M.~Bardeen \textit{et al.},
``Rotating Black Holes: Locally Nonrotating Frames, Energy Extraction, and Scalar Synchrotron Radiation,''
Astrophys. J. \textbf{178}, 347-370 (1972)
doi:10.1086/151796.

\bibitem{Teo:2020sey}
E.~Teo,
``Spherical orbits around a Kerr black hole,''
Gen. Rel. Grav. \textbf{53}, no.1, 10 (2021)
doi:10.1007/s10714-020-02782-z
[arXiv:2007.04022 [gr-qc]].


\bibitem{2024ApJ...966..226K}
O.~Kop\'{a}\v{c}ek and V.~Karas,
``On Innermost Stable Spherical Orbits near a Rotating Black Hole: A Numerical Study of the Particle Motion near the Plunging Region,''
Astrophys. J. \textbf{966}, no.~2, 226 (2024)
doi:10.3847/1538-4357/ad3932
[arXiv:2404.04501 [astro-ph.HE]].


\bibitem{Carter:1968rr}
B.~Carter,
``Global structure of the Kerr family of gravitational fields,''
Phys. Rev. \textbf{174}, 1559-1571 (1968)
doi:10.1103/PhysRev.174.1559



\bibitem{Sun:2021oxg}
W.~Sun, Y.~Wang, F.~Liu and X.~Wu,
``Applying explicit symplectic integrator to study chaos of charged particles around magnetized Kerr black hole,''
Eur. Phys. J. C \textbf{81}, no.9, 785 (2021)
doi:10.1140/epjc/s10052-021-09579-7
[arXiv:2109.02295 [gr-qc]].


\bibitem{Cao:2022bvu}
W.~Cao, W.~Liu and X.~Wu,
``Integrability of Kerr-Newman spacetime with cloud strings, quintessence and electromagnetic field,''
Phys. Rev. D \textbf{105}, no.12, 124039 (2022)
doi:10.1103/PhysRevD.105.124039
[arXiv:2206.09518 [gr-qc]].



\bibitem{Tancredi:2001}
G.~Tancredi, A.~S$\acute{a}$nchez, and F.~Roig,
``A Comparison Between Methods to Compute Lyapunov Exponents,''
Astron. J. \textbf{121}, no.2, 1171-1179 (2001)
doi:10.1086/318732.



\bibitem{Wu:2003pe}
X.~Wu and T.~y.~Huang,
``Computation of Lyapunov exponents in general relativity,''
Phys. Lett. A \textbf{313}, 77-81 (2003)
doi:10.1016/S0375-9601(03)00720-5
[arXiv:gr-qc/0302118 [gr-qc]].


\bibitem{Wu:2006rx}
X.~Wu, T.~Y.~Huang and H.~Zhang,
``Lyapunov indices with two nearby trajectories in a curved spacetime,''
Phys. Rev. D \textbf{74}, 083001 (2006)
doi:10.1103/PhysRevD.74.083001
[arXiv:1006.5251 [gr-qc]].


\bibitem{Cardoso:2008bp}
V.~Cardoso, A.~S.~Miranda, E.~Berti, H.~Witek and V.~T.~Zanchin,
``Geodesic stability, Lyapunov exponents and quasinormal modes,''
Phys. Rev. D \textbf{79}, no.6, 064016 (2009)
doi:10.1103/PhysRevD.79.064016
[arXiv:0812.1806 [hep-th]].


\bibitem{Ma:2014aha}
D.~Z.~Ma, J.~P.~Wu and J.~Zhang,
``Chaos from the ring string in a Gauss-Bonnet black hole in AdS5 space,''
Phys. Rev. D \textbf{89}, no.8, 086011 (2014)
doi:10.1103/PhysRevD.89.086011
[arXiv:1405.3563 [hep-th]].




\bibitem{Lei:2020clg}
Y.~Q.~Lei, X.~H.~Ge and C.~Ran,
``Chaos of particle motion near a black hole with quasitopological electromagnetism,''
Phys. Rev. D \textbf{104}, no.4, 046020 (2021)
doi:10.1103/PhysRevD.104.046020
[arXiv:2008.01384 [hep-th]].


\bibitem{Kostaros:2021usv}
K.~Kostaros and G.~Pappas,
``Chaotic photon orbits and shadows of a non-Kerr object described by the Hartle{\textendash}Thorne spacetime,''
Class. Quant. Grav. \textbf{39}, no.13, 134001 (2022)
doi:10.1088/1361-6382/ac7028
[arXiv:2111.09367 [gr-qc]].
\end{thebibliography}
\end{document}